\documentclass[12pt,preprint]{aastex} 

\def\gtorder{\mathrel{\raise.3ex\hbox{$>$}\mkern-14mu
             \lower0.6ex\hbox{$\sim$}}} 
\def\ltsima{$\; \buildrel < \over \sim \;$}
\def\simlt{\lower.5ex\hbox{\ltsima}}
\def\gtsima{$\; \buildrel > \over \sim \;$}
\def\simgt{\lower.5ex\hbox{\gtsima}} 

\begin{document} 


\title{Supernovae in Low-Redshift Galaxy Clusters:\\
Observations by the Wise Observatory Optical Transient Search (WOOTS)} 


\author{Avishay Gal-Yam}
\affil{Astrophysics Group, Faculty of Physics, The Weizmann
Institute of Science, Rehovot 76100, Israel}
\email{avishay.gal-yam@weizmann.ac.il}

\author{Dan Maoz}
\affil{School of Physics and Astronomy, Tel Aviv University, Tel Aviv 69978, Israel}

\author{Puragra Guhathakurta}
\affil{UCO/Lick Observatory, Department of Astronomy \& Astrophysics, University of California, 1156 High
Street, Santa Cruz, CA 95064, USA}

\and

\author{Alexei V. Filippenko}
\affil{Department of Astronomy, University of California, Berkeley, CA 94720-3411, USA}



\begin{abstract} 

We describe the Wise Observatory Optical Transient Search (WOOTS), a
survey for supernovae (SNe) and other variable and transient objects in the
fields of redshift 0.06--0.2 Abell galaxy clusters. We present the survey
design and data-analysis procedures, and our object detection and follow-up
strategies. We have obtained follow-up spectroscopy for all viable SN
candidates, and present the resulting SN sample here.
Out of the 12 SNe we have discovered, seven are associated with our target
clusters while five are foreground or background field events. All but one
of the SNe (a foreground field event) are Type Ia SNe.
Our non-cluster
SN sample is uniquely complete, since all SN candidates have been
either spectroscopically confirmed or ruled out. This allows us to 
estimate that flux-limited surveys similar to WOOTS would
be dominated ($\sim80\%$) by SNe~Ia. Our spectroscopic follow-up
observations also elucidate the difficulty in distinguishing active
galactic nuclei from SNe.
In separate papers we use the WOOTS sample to derive the SN rate
in clusters for this redshift range, and to measure the fraction of
intergalactic cluster SNe. We also briefly
report here on some quasars and asteroids discovered by WOOTS.   



\end{abstract} 






\section{Introduction} 

Searches for extragalactic variable or transient astronomical phenomena
are almost as old as modern optical astronomy. Well-known examples are the
searches for extragalactic variable stars by Edwin Hubble
(e.g., Hubble \& Humason 1931) and his successors, and the pioneering searches
for supernovae (SNe) by Fritz Zwicky (e.g., Zwicky, Humason, \& Gates 1960).
For many decades, these searches were mostly targeted at specific
nearby (redshift $z < 0.1$) galaxies or clusters. 
In the late 1980s, Norgaard-Nielsen et al. (1989) undertook a
survey for SNe in more distant galaxy clusters, resulting
in the discovery of a Type Ia supernova (SN~Ia) in the cluster AC 118 ($z=0.31$). This
breakthrough achievement illustrated the potential of such surveys,
though the observational costs of this project were considerable.
Perlmutter et al. (1990, 1995)
used the emerging new technologies of wide-field CCD imaging and
image subtraction to carry out the first efficient surveys for
high-redshift SNe in clusters and in the field. These methods, employed
by large international collaborations (e.g., Perlmutter et al. 1997, Schmidt et
al. 1998), have resulted in the discovery of hundreds of high-$z$
SNe~Ia. Analysis of these objects indicates that the expansion of the
Universe is currently accelerating (Riess et al. 1998, 2004, 2007;
Perlmutter et al. 1999; Tonry et al. 2003; Knop et al. 2003;
Astier et al. 2006; Wood-Vasey et al. 2007); see Filippenko (2005b)
for a review.   

SNe are not the only variable or
transient objects that are luminous enough to be observed in distant galaxies.
Other types of objects include variable active
galactic nuclei (AGNs; e.g., Hawkins \& Veron 1993; Gal-Yam et al. 2002;
Sarajedini et al. 2006) of various sorts, and
optical transients associated with cosmological gamma-ray bursts (GRBs;
e.g. van Paradijs et al. 1997).  
We have undertaken a survey for SNe and other transient or variable sources
in the fields of galaxy clusters, the Wise Observatory Optical Transient
Search (WOOTS; Gal-Yam \& Maoz 2000a,b; Gal-Yam et al. 2005). Our main
motivation is to study SNe in galaxy clusters. In particular, measurement
of the cluster SN Ia rate has interesting implications for SN Ia
progenitors and the origin of metals in the intracluster medium (ICM;
e.g., Gal-Yam, Maoz, \& Sharon 2002; Maoz \& Gal-Yam 2004). Another
objective is to discover and quantify the occurrence of
intergalactic SNe (Gal-Yam et al. 2003a).  

Here, we describe the
technical design of this survey, the data-analysis methods (\S~2), and
the SNe we discovered (\S~3.1), along with a brief report
on other types of objects (\S~3.2). In a companion paper (Sharon et al. 2007;
Paper II) we derive from our SN sample a
measurement of the rate of SNe in galaxy clusters.
The discovery of intergalactic SNe by WOOTS was reported by Gal-Yam et
al. (2003a).  

\section{Technical Design} 

\subsection{The Cluster Sample} 

The cluster sample we surveyed was drawn from the only large galaxy-cluster
catalog available at the time, by Abell, Corwin, \& Olowin (1989; hereafter
ACO). This catalog is based on selection of galaxy overdensities from
photographic plates. It is thus prone to projection problems, particularly to
superposition of several poor clusters or groups along the same line of
sight, appearing as single richer systems.
However, the catalog was shown to be largely uncontaminated by such
superpositions
in the redshift range considered for our study ($z < 0.2$; Lucey
1983). At the time 
WOOTS was initiated, no large, X-ray-selected cluster catalogs were
available. 

The clusters were selected using the following criteria. First, all ACO clusters
with measured redshifts $0.06 < z < 0.2$ in the compilation of Struble \& Rood
(1991) were considered. This redshift range was chosen so as to include
clusters that are generally more distant than those studied by previous search
programs (e.g., Reiss et al. 1998; $z<0.08$), while still sufficiently nearby
that SNe in the cluster galaxies are detectable with the $1~$m telescope at
Wise Observatory.
To ensure proper visibility from Wise, we limited the sample
to northern-hemisphere clusters with declinations $\delta > 0$. Seeking to
increase the number of galaxies per field, we selected only rich clusters (Abell
richness class $R \ge 1$) with an Abell galaxy count of $N > 65$. We also
required that the radius of the clusters, as estimated by Leir
\& van den Bergh (1977), be smaller than $20'$, in order for most of
the cluster galaxies to fall within the $12' \times 12'$ field of view of the Wise
Observatory Tektronix CCD imager. Applying these cuts to the ACO catalog, we ended up
with 161 galaxy clusters. During the survey runs, we observed two additional
clusters that had similar properties and were known gravitational lenses.
The final cluster sample, including 163 entries, is listed in
Table 2.1\footnote{Due to technical difficulties in data acquisition and
recovery, 23 clusters were excluded from the analysis in Paper II, and
thus the cluster sample reported there includes only 140 clusters.}. 

The cluster sample defined above is not uniformly
distributed on the sky, with most of the targets at relatively high
Galactic latitudes. In order to optimize the use of allocated telescope
time, we supplemented our cluster target list with a secondary list of
empty fields in the vicinity of the clusters residing in sparsely
populated sky areas (Abell numbers 24--913; 2244--2694). This secondary list
of targets was imaged when additional telescope time was available,
after all suitable cluster targets had been imaged. These additional data were 
searched for SNe in the same manner as the cluster fields (leading,
e.g., to the discovery of SN 1999fd).



\begin{table}
\caption {WOOTS Cluster Sample} \label {cluster table}
\vspace{0.2cm}
\begin{centering}
\begin{minipage}{140mm}
\begin{scriptsize}
\begin{tabular}{cccccccc}
\hline
Target & Abell  & RA       & Declination &  Redshift & Richness & Radius   & Galaxy \\
Number & Number & (B1950)  & (B1950)     &  $z$      & Class    & (arcmin) & Count  \\   
\hline 
  1    &    24  & 00$^h$19$^m$54$^s$  & $23^\circ01'00''$    &  0.134    &     2    &   20     &   127  \\              
  2    &    31  & 00$^h$24$^m$30$^s$  & $22^\circ21'00''$    &  0.160    &     2    &   10     &    90  \\  
  3    &    79  & 00$^h$38$^m$00$^s$  & $17^\circ52'00''$    &  0.093    &     1    &    8     &    73  \\
  4    &    84  & 00$^h$39$^m$12$^s$  & $21^\circ08'00''$    &  0.103    &     1    &    8     &    76  \\
  5    &    98  & 00$^h$43$^m$48$^s$  & $20^\circ13'00''$    &  0.104    &     3    &   20     &   185  \\ 
  6    &   115  & 00$^h$53$^m$18$^s$  & $26^\circ03'00''$    &  0.197    &     3    &   13     &   174  \\
  7    &   125  & 00$^h$57$^m$30$^s$  & $14^\circ00'00''$    &  0.188    &     1    &   12     &    66  \\
  8    &   136  & 01$^h$01$^m$24$^s$  & $24^\circ48'00''$    &  0.157    &     2    &   12     &    99  \\       
  9    &   175  & 01$^h$16$^m$54$^s$  & $14^\circ37'00''$    &  0.129    &     2    &   20     &    84  \\          
10    &   234  & 01$^h$38$^m$18$^s$  & $18^\circ40'00''$    &  0.173    &     1    &   11     &    76  \\        
11    &   279  & 01$^h$53$^m$48$^s$  & $00^\circ49'00''$    &  0.080    &     1    &   11     &    70  \\ 
12    &   293  & 01$^h$59$^m$24$^s$  & $03^\circ32'00''$    &  0.163    &     2    &   15     &    87  \\        
13    &   403  & 02$^h$56$^m$36$^s$  & $03^\circ18'00''$    &  0.103    &     2    &   13     &   100  \\  
14    &   410  & 03$^h$01$^m$18$^s$  & $03^\circ36'00''$    &  0.090    &     1    &   14     &    70  \\         
15    & Ms0440 & 04$^h$40$^m$31$^s$  & $02^\circ04'37''$    &  0.190    &     -    &    -     &     -  \\        
16    &   508  & 04$^h$43$^m$18$^s$  & $01^\circ55'00''$    &  0.148    &     2    &   12     &    85  \\
17    &   509  & 04$^h$45$^m$06$^s$  & $02^\circ12'00''$    &  0.084    &     1    &   15     &    72  \\
18    &   562  & 06$^h$46$^m$30$^s$  & $69^\circ20'00''$    &  0.110    &     1    &   16     &    70  \\       
19    &   588  & 07$^h$33$^m$36$^s$  & $70^\circ04'00''$    &  0.160    &     1    &   12     &    78  \\        
20    &   629  & 08$^h$10$^m$18$^s$  & $66^\circ35'00''$    &  0.138    &     1    &    8     &    78  \\       
21    &   655  & 08$^h$21$^m$48$^s$  & $47^\circ17'00''$    &  0.124    &     3    &   14     &   142  \\  
22    &   665  & 08$^h$26$^m$12$^s$  & $66^\circ03'00''$    &  0.182    &     5    &   11     &   321  \\         
23    &   750  & 09$^h$06$^m$24$^s$  & $11^\circ14'00''$    &  0.162    &     3    &   18     &   142  \\        
24    &   795  & 09$^h$21$^m$18$^s$  & $14^\circ23'00''$    &  0.136    &     3    &   13     &   151  \\        
25    &   801  & 09$^h$25$^m$12$^s$  & $20^\circ47'00''$    &  0.192    &     2    &    9     &    81  \\       
26    &   873  & 09$^h$48$^m$24$^s$  & $71^\circ32'00''$    &  0.182    &     3    &   20     &   133  \\       
27    &   913  & 09$^h$59$^m$48$^s$  & $20^\circ43'00''$    &  0.168    &     1    &    8     &    65  \\       
28    &   914  & 10$^h$04$^m$18$^s$  & $71^\circ29'00''$    &  0.195    &     2    &   12     &   114  \\       
29    &   917  & 10$^h$04$^m$12$^s$  & $62^\circ45'00''$    &  0.132    &     1    &   12     &    66  \\       
30    &   918  & 10$^h$06$^m$06$^s$  & $73^\circ59'00''$    &  0.167    &     1    &   20     &    69  \\       
31    &   922  & 10$^h$06$^m$18$^s$  & $71^\circ16'00''$    &  0.190    &     2    &   10     &    85  \\       
32    &   924  & 10$^h$04$^m$06$^s$  & $35^\circ54'00''$    &  0.099    &     1    &   10     &    75  \\       
33    &   947  & 10$^h$11$^m$36$^s$  & $63^\circ19'00''$    &  0.177    &     1    &   16     &    77  \\       
34    &   960  & 10$^h$15$^m$06$^s$  & $66^\circ28'00''$    &  0.129    &     2    &   11     &   117  \\       
35    &   968  & 10$^h$17$^m$24$^s$  & $68^\circ31'00''$    &  0.195    &     2    &    9     &   119  \\       
36    &   975  & 10$^h$19$^m$06$^s$  & $65^\circ42'00''$    &  0.116    &     2    &   13     &    84  \\       
38    &  1025  & 10$^h$28$^m$12$^s$  & $63^\circ06'00''$    &  0.151    &     2    &   15     &    87  \\       
39    &  1029  & 10$^h$31$^m$00$^s$  & $77^\circ35'00''$    &  0.126    &     2    &   15     &    81  \\       
40    &  1046  & 10$^h$34$^m$00$^s$  & $68^\circ13'00''$    &  0.190    &     2    &   10     &   108  \\       
41    &  1061  & 10$^h$37$^m$18$^s$  & $67^\circ28'00''$    &  0.189    &     2    &    7     &    99  \\       
42    &  1066  & 10$^h$36$^m$48$^s$  & $05^\circ26'00''$    &  0.070    &     1    &   16     &    68  \\       
43    &  1073  & 10$^h$39$^m$36$^s$  & $36^\circ54'00''$    &  0.139    &     2    &   18     &    82  \\       
44    &  1081  & 10$^h$42$^m$00$^s$  & $35^\circ50'00''$    &  0.159    &     2    &   15     &    83  \\       
45    &  1123  & 10$^h$52$^m$54$^s$  & $75^\circ47'00''$    &  0.123    &     2    &   15     &   108  \\       
46    &  1132  & 10$^h$55$^m$18$^s$  & $57^\circ03'00''$    &  0.136    &     1    &   15     &    74  \\       
47    &  1170  & 11$^h$04$^m$54$^s$  & $08^\circ17'00''$    &  0.162    &     2    &   13     &   104  \\       
48    &  1190  & 11$^h$09$^m$00$^s$  & $41^\circ07'00''$    &  0.079    &     2    &   20     &    87  \\       
49    &  1201  & 11$^h$10$^m$24$^s$  & $13^\circ42'00''$    &  0.169    &     2    &   15     &   103  \\       
50    &  1207  & 11$^h$12$^m$36$^s$  & $67^\circ58'00''$    &  0.135    &     1    &   11     &    72  \\       
51    &  1227  & 11$^h$18$^m$48$^s$  & $48^\circ18'00''$    &  0.112    &     2    &   13     &   112  \\       
52    &  1232  & 11$^h$19$^m$30$^s$  & $18^\circ10'00''$    &  0.168    &     1    &   12     &    70  \\       
53    &  1234  & 11$^h$19$^m$48$^s$  & $21^\circ40'00''$    &  0.166    &     2    &   11     &    88  \\       
54    &  1235  & 11$^h$20$^m$18$^s$  & $19^\circ54'00''$    &  0.104    &     2    &   20     &   122  \\       
55    &  1249  & 11$^h$22$^m$24$^s$  & $68^\circ18'00''$    &  0.156    &     1    &   10     &    77  \\       
\hline
\end{tabular}
\end{scriptsize} 
\end{minipage}
\end{centering}
\end{table} 

\begin{table}
\vspace{0.2cm}
\begin{centering}
\begin{minipage}{140mm}
\begin{scriptsize}
\begin{tabular}{cccccccc}
\hline
Target & Abell  & Right     & Declination &  Redshift & Richness & Radius   & Galaxy \\
Number & Number & Ascension & (B1950)     &  $z$      & Class    & (arcmin) & Count  \\   
\hline
56    &  1255  & 11$^h$24$^m$24$^s$  & $75^\circ45'00''$    &  0.166    &     1    &   13     &    77  \\       
57    &  1264  & 11$^h$24$^m$36$^s$  & $17^\circ25'00''$    &  0.127    &     2    &   20     &    99  \\       
58    &  1278  & 11$^h$27$^m$36$^s$  & $20^\circ45'00''$    &  0.129    &     3    &   15     &   151  \\       
59    &  1302  & 11$^h$30$^m$30$^s$  & $66^\circ41'00''$    &  0.116    &     2    &   15     &    85  \\       
60    &  1307  & 11$^h$30$^m$12$^s$  & $14^\circ48'00''$    &  0.083    &     1    &   13     &    71  \\       
61    &  1345  & 11$^h$38$^m$36$^s$  & $10^\circ58'00''$    &  0.109    &     1    &   13     &    71  \\       
62    &  1349  & 11$^h$39$^m$24$^s$  & $55^\circ38'00''$    &  0.136    &     1    &   17     &    66  \\       
63    &  1356  & 11$^h$39$^m$54$^s$  & $10^\circ43'00''$    &  0.070    &     1    &   10     &    77  \\       
64    &  1360  & 11$^h$40$^m$30$^s$  & $11^\circ18'00''$    &  0.154    &     1    &    9     &    66  \\       
65    &  1372  & 11$^h$42$^m$54$^s$  & $11^\circ48'00''$    &  0.113    &     1    &   10     &    70  \\       
66    &  1381  & 11$^h$45$^m$42$^s$  & $75^\circ30'00''$    &  0.117    &     2    &   15     &    92  \\       
67    &  1401  & 11$^h$49$^m$30$^s$  & $37^\circ33'00''$    &  0.165    &     3    &   17     &   153  \\       
68    &  1406  & 11$^h$50$^m$36$^s$  & $68^\circ10'00''$    &  0.118    &     1    &   10     &    69  \\       
69    &  1408  & 11$^h$51$^m$12$^s$  & $15^\circ40'00''$    &  0.110    &     1    &   15     &    72  \\       
70    &  1412  & 11$^h$53$^m$06$^s$  & $73^\circ45'00''$    &  0.084    &     2    &   13     &    86  \\         
71    &  1413  & 11$^h$52$^m$48$^s$  & $23^\circ39'00''$    &  0.143    &     3    &   16     &   196  \\       
72    &  1415  & 11$^h$53$^m$12$^s$  & $58^\circ09'00''$    &  0.159    &     1    &   10     &    66  \\       
73    &  1421  & 11$^h$54$^m$24$^s$  & $68^\circ15'00''$    &  0.119    &     1    &   10     &    65  \\       
74    &  1437  & 11$^h$57$^m$54$^s$  & $03^\circ37'00''$    &  0.134    &     3    &   13     &   154  \\       
75    &  1445  & 11$^h$59$^m$12$^s$  & $00^\circ07'00''$    &  0.169    &     2    &   13     &    81  \\       
76    &  1446  & 11$^h$59$^m$18$^s$  & $58^\circ18'00''$    &  0.104    &     2    &   17     &    85  \\       
77    &  1470  & 12$^h$04$^m$24$^s$  & $71^\circ55'00''$    &  0.192    &     2    &   12     &    93  \\       
78    &  1474  & 12$^h$05$^m$24$^s$  & $15^\circ14'00''$    &  0.079    &     1    &   18     &    70  \\       
79    &  1477  & 12$^h$06$^m$24$^s$  & $64^\circ21'00''$    &  0.111    &     1    &    5     &    71  \\       
80    &  1495  & 12$^h$10$^m$24$^s$  & $29^\circ31'00''$    &  0.143    &     2    &   15     &   123  \\       
81    &  1497  & 12$^h$11$^m$36$^s$  & $26^\circ56'00''$    &  0.167    &     2    &   14     &   101  \\       
82    &  1504  & 12$^h$12$^m$48$^s$  & $27^\circ48'00''$    &  0.184    &     2    &   14     &    98  \\       
83    &  1514  & 12$^h$15$^m$24$^s$  & $20^\circ56'00''$    &  0.199    &     3    &   18     &   132  \\         
84    &  1524  & 12$^h$19$^m$12$^s$  & $08^\circ07'00''$    &  0.137    &     2    &    6     &   103  \\       
85    &  1528  & 12$^h$20$^m$30$^s$  & $59^\circ11'00''$    &  0.154    &     1    &   12     &    75  \\       
86    &  1539  & 12$^h$24$^m$00$^s$  & $62^\circ50'00''$    &  0.171    &     2    &   12     &    96  \\       
87    &  1548  & 12$^h$26$^m$30$^s$  & $19^\circ42'00''$    &  0.161    &     3    &   10     &   155  \\       
88    &  1552  & 12$^h$27$^m$18$^s$  & $12^\circ01'00''$    &  0.084    &     1    &   11     &    75  \\       
89    &  1553  & 12$^h$28$^m$18$^s$  & $10^\circ51'00''$    &  0.165    &     2    &   10     &   100  \\       
90    &  1562  & 12$^h$31$^m$48$^s$  & $41^\circ27'00''$    &  0.190    &     1    &   11     &    77  \\       
91    &  1566  & 12$^h$32$^m$48$^s$  & $64^\circ39'00''$    &  0.101    &     2    &   10     &    91  \\       
92    &  1607  & 12$^h$41$^m$00$^s$  & $76^\circ25'00''$    &  0.136    &     2    &   12     &    82  \\       
93    &  1617  & 12$^h$45$^m$24$^s$  & $59^\circ28'00''$    &  0.152    &     3    &    9     &   139  \\       
94    &  1632  & 12$^h$50$^m$36$^s$  & $29^\circ05'00''$    &  0.196    &     2    &   10     &    80  \\         
95    &  1661  & 12$^h$59$^m$24$^s$  & $29^\circ21'00''$    &  0.167    &     2    &   10     &    97  \\         
96    &  1667  & 13$^h$00$^m$24$^s$  & $32^\circ05'00''$    &  0.165    &     2    &   10     &    98  \\         
97    &  1674  & 13$^h$01$^m$42$^s$  & $67^\circ46'00''$    &  0.107    &     3    &   12     &   165  \\         
98    &  1677  & 13$^h$03$^m$30$^s$  & $31^\circ10'00''$    &  0.183    &     2    &    9     &   112  \\         
99    &  1678  & 13$^h$03$^m$00$^s$  & $62^\circ31'00''$    &  0.170    &     1    &    8     &    78  \\         
100    &  1679  & 13$^h$04$^m$12$^s$  & $32^\circ04'00''$    &  0.170    &     2    &   10     &   115  \\         
101    &  1681  & 13$^h$03$^m$24$^s$  & $72^\circ08'00''$    &  0.091    &     1    &   10     &    79  \\         
102    &  1697  & 13$^h$10$^m$42$^s$  & $46^\circ31'00''$    &  0.183    &     2    &    9     &    84  \\         
103    &  1731  & 13$^h$20$^m$48$^s$  & $58^\circ26'00''$    &  0.193    &     2    &   15     &    92  \\         
104    &  1738  & 13$^h$23$^m$12$^s$  & $57^\circ52'00''$    &  0.115    &     2    &   18     &    85  \\         
105    &  1741  & 13$^h$22$^m$42$^s$  & $71^\circ44'00''$    &  0.075    &     1    &   15     &    73  \\        
106    &  1759  & 13$^h$31$^m$36$^s$  & $20^\circ30'00''$    &  0.168    &     3    &   10     &   132  \\        
107    &  1760  & 13$^h$31$^m$42$^s$  & $20^\circ28'00''$    &  0.171    &     3    &   11     &   168  \\        
108    &  1763  & 13$^h$33$^m$06$^s$  & $41^\circ13'00''$    &  0.187    &     3    &   10     &   152  \\        
109    &  1767  & 13$^h$34$^m$12$^s$  & $59^\circ28'00''$    &  0.070    &     1    &   18     &    65  \\        
110    &  1773  & 13$^h$39$^m$36$^s$  & $02^\circ30'00''$    &  0.078    &     1    &   10     &    66  \\        
\hline
\end{tabular}
\end{scriptsize} 
\end{minipage}
\end{centering}
\end{table} 

\begin{table}
\vspace{0.2cm}
\begin{centering}
\begin{minipage}{140mm}
\begin{scriptsize}
\begin{tabular}{cccccccc}
\hline
Target & Abell  & Right     & Declination &  Redshift & Richness & Radius   & Galaxy \\
Number & Number & Ascension & (B1950)     &  $z$      & Class    & (arcmin) & Count  \\   
\hline
111    &  1774  & 13$^h$39$^m$00$^s$  & $40^\circ16'00''$    &  0.169    &     2    &   10     &    81  \\        
112    &  1780  & 13$^h$42$^m$06$^s$  & $03^\circ08'00''$    &  0.079    &     1    &   16     &    71  \\        
113    &  1795  & 13$^h$46$^m$42$^s$  & $26^\circ50'00''$    &  0.062    &     2    &   18     &   115  \\        
114    &  1827  & 13$^h$55$^m$54$^s$  & $21^\circ57'00''$    &  0.067    &     1    &   16     &    68  \\        
115    &  1852  & 14$^h$01$^m$36$^s$  & $16^\circ00'00''$    &  0.181    &     1    &    9     &    77  \\        
116    &  1880  & 14$^h$10$^m$54$^s$  & $22^\circ38'00''$    &  0.141    &     1    &   15     &    67  \\         
117    &  1889  & 14$^h$14$^m$30$^s$  & $30^\circ57'00''$    &  0.186    &     2    &   11     &   112  \\         
118    &  1892  & 14$^h$10$^m$24$^s$  & $78^\circ58'00''$    &  0.091    &     1    &    7     &    79  \\         
119    &  1904  & 14$^h$20$^m$18$^s$  & $48^\circ47'00''$    &  0.071    &     2    &   20     &    83  \\         
120    &  1909  & 14$^h$21$^m$48$^s$  & $25^\circ09'00''$    &  0.146    &     1    &   16     &    79  \\         
121    &  1911  & 14$^h$22$^m$24$^s$  & $39^\circ11'00''$    &  0.191    &     2    &    8     &    80  \\         
122    &  1914  & 14$^h$24$^m$00$^s$  & $38^\circ03'00''$    &  0.171    &     2    &   13     &   105  \\         
123    &  1918  & 14$^h$23$^m$54$^s$  & $63^\circ23'00''$    &  0.140    &     3    &   10     &   142  \\         
124    &  1920  & 14$^h$25$^m$42$^s$  & $56^\circ00'00''$    &  0.131    &     2    &   12     &   103  \\         
125    &  1926  & 14$^h$28$^m$18$^s$  & $24^\circ52'00''$    &  0.132    &     2    &   15     &   112  \\        
126    &  1936  & 14$^h$32$^m$54$^s$  & $55^\circ02'00''$    &  0.139    &     1    &    8     &    69  \\        
127    &  1937  & 14$^h$33$^m$00$^s$  & $58^\circ29'00''$    &  0.139    &     2    &    8     &    99  \\        
128    &  1940  & 14$^h$33$^m$54$^s$  & $55^\circ22'00''$    &  0.140    &     3    &   13     &   130  \\        
129    &  1954  & 14$^h$39$^m$54$^s$  & $28^\circ44'00''$    &  0.181    &     2    &    8     &   120  \\        
130    &  1966  & 14$^h$42$^m$48$^s$  & $59^\circ06'00''$    &  0.151    &     2    &    5     &   104  \\        
131    &  1979  & 14$^h$48$^m$54$^s$  & $31^\circ29'00''$    &  0.169    &     2    &    9     &   108  \\        
132    &  1984  & 14$^h$50$^m$12$^s$  & $28^\circ09'00''$    &  0.124    &     2    &   11     &    93  \\       
133    &  1986  & 14$^h$50$^m$54$^s$  & $22^\circ07'00''$    &  0.118    &     1    &   10     &    67  \\        
134    &  1990  & 14$^h$51$^m$36$^s$  & $28^\circ17'00''$    &  0.127    &     3    &   11     &   140  \\        
135    &  1999  & 14$^h$52$^m$36$^s$  & $54^\circ31'00''$    &  0.103    &     1    &   16     &    68  \\        
136    &  2005  & 14$^h$56$^m$36$^s$  & $28^\circ01'00''$    &  0.126    &     2    &   11     &   105  \\        
137    &  2008  & 14$^h$57$^m$48$^s$  & $23^\circ20'00''$    &  0.181    &     2    &    9     &    93  \\        
138    &  2009  & 14$^h$58$^m$00$^s$  & $21^\circ34'00''$    &  0.153    &     1    &   10     &    66  \\        
139    &  2029  & 15$^h$08$^m$30$^s$  & $05^\circ57'00''$    &  0.077    &     2    &   20     &    82  \\         
140    &  2048  & 15$^h$12$^m$48$^s$  & $04^\circ34'00''$    &  0.094    &     1    &   13     &    75  \\        
141    &  2053  & 15$^h$14$^m$42$^s$  & $00^\circ30'00''$    &  0.113    &     1    &   10     &    75  \\        
142    &  2061  & 15$^h$19$^m$12$^s$  & $30^\circ50'00''$    &  0.078    &     1    &   20     &    71  \\        
143    &  2062  & 15$^h$19$^m$18$^s$  & $32^\circ16'00''$    &  0.112    &     1    &   15     &    69  \\        
144    &  2069  & 15$^h$21$^m$54$^s$  & $30^\circ04'00''$    &  0.116    &     2    &   13     &    97  \\        
145    &  2089  & 15$^h$30$^m$36$^s$  & $28^\circ11'00''$    &  0.073    &     1    &   20     &    70  \\        
146    &  2100  & 15$^h$34$^m$30$^s$  & $37^\circ48'00''$    &  0.153    &     3    &   10     &   138  \\        
147    &  2122  & 15$^h$42$^m$36$^s$  & $36^\circ17'00''$    &  0.065    &     1    &    8     &    68  \\         
148    &  2172  & 16$^h$15$^m$06$^s$  & $42^\circ31'00''$    &  0.139    &     1    &    8     &    69  \\          
149    &  2198  & 16$^h$26$^m$30$^s$  & $43^\circ56'00''$    &  0.080    &     2    &    6     &    85  \\         
150    &  2213  & 16$^h$34$^m$54$^s$  & $41^\circ23'00''$    &  0.160    &     1    &    8     &    75  \\         
151    &  2218  & 16$^h$35$^m$42$^s$  & $66^\circ19'00''$    &  0.171    &     4    &   10     &   214  \\          
152    &  2235  & 16$^h$53$^m$18$^s$  & $40^\circ06'00''$    &  0.151    &     1    &    8     &    73  \\         
153    &  2240  & 16$^h$54$^m$00$^s$  & $66^\circ49'00''$    &  0.138    &     3    &   14     &   165  \\         
154    &  2244  & 17$^h$00$^m$54$^s$  & $34^\circ07'00''$    &  0.097    &     2    &   16     &    89  \\         
155    &  2255  & 17$^h$12$^m$12$^s$  & $64^\circ09'00''$    &  0.081    &     2    &   20     &   102  \\         
156    &  2283  & 17$^h$44$^m$54$^s$  & $69^\circ40'00''$    &  0.183    &     1    &   10     &    65  \\         
157    &  2355  & 21$^h$32$^m$48$^s$  & $01^\circ10'00''$    &  0.124    &     2    &    8     &   112  \\           
158    &  2356  & 21$^h$33$^m$12$^s$  & $00^\circ06'00''$    &  0.116    &     2    &   16     &    89  \\       
159    &  2390  & 21$^h$51$^m$12$^s$  & $17^\circ26'00''$    &  0.231    &     -    &    -     &     -  \\       
160    &  2471  & 22$^h$39$^m$18$^s$  & $07^\circ00'00''$    &  0.108    &     2    &   10     &    92  \\       
161    &  2616  & 23$^h$30$^m$42$^s$  & $05^\circ20'00''$    &  0.183    &     2    &    7     &    94  \\       
162    &  2623  & 23$^h$32$^m$30$^s$  & $05^\circ20'00''$    &  0.178    &     3    &   11     &   142  \\       
163    &  2694  & 00$^h$00$^m$00$^s$  & $08^\circ09'00''$    &  0.096    &     3    &   13     &   132  \\       
\hline
\end{tabular}
\end{scriptsize} 
\end{minipage}
\end{centering}
\end{table} 

\subsection{Observations, Reduction, and Analysis}  

\subsubsection{Observing Strategy} 

The main objective of WOOTS, to detect SNe out to $z=0.2$,
is challenging for an imaging survey with a $1~$m telescope. We used
$600$~s unfiltered images which, under good
conditions at Wise, reach a point-source detection limit equivalent to
$R\approx22$ mag ($3\sigma$; tied to photometric zeropoints
derived from USNO-A1 catalog magnitudes; Monet et al. 1998).
Unfiltered images obtained with the Wise cameras typically reach flux
limits that are a factor of $\sim 2$ fainter than broad-band $R$ or
$V$ imaging. This allowed us to detect bright SNe~Ia (with a peak
$R$-band magnitude of $\sim 20.5$ at $z=0.2$), as well as somewhat fainter events.
We were allocated $3-5$ nights per month on the Wise $1~$m, and were able
to observe approximately 40 cluster fields per night.
In Paper II we discuss in some detail the photometric calibration of WOOTS unfiltered
magnitudes using Sloan Digital Sky Survey (SDSS) data that have subsequently
become available. 

We split the $600$~s observation of each field into $3 \times 200$ s frames that
were obtained within an interval of $\sim 30$ min. The splits allowed us to reject
cosmic rays and chip defects, while the time staggering is important for locating
Solar System objects that may appear to be stationary transient sources if one
obtains only a single image.  

Between Oct. 1997 and Dec. 1999 we used the Tektronix CCD imager,
which is a thinned, back-illuminated chip with $1024 \times 1024$ pixels,
each $0.''7$ on a side, resulting in a $12' \times 12'$ field of view.
In Mar. 2000, we commissioned and began using a new, blue-sensitized SITe
CCD camera having $2048 \times 4096$ pixels, each $0.''4$ on a side, with a
total field of view of $\sim 15' \times 30'$. However,
initial technical problems with the camera electronics
and operating system, as well as increased problems in data analysis,
limited the results from these observations.
Only one supernova, SN 2001al, was discovered with the SITe CCD.
In Paper~II, we analyze only the 1997--1999 observations taken with the
Tektronix CCD.  

\subsubsection{Data Reduction} 

Bias-subtracted WOOTS images were reduced in the following manner. 
We constructed ``superflat'' calibration frames from
the actual science data by median
combining the stacks of images obtained each night. This method worked well
and was far superior to using twilight-sky images (``sky flats'') or
images of an illuminated white calibration surface (``dome flats''), probably due to
the difference in color between the calibration illuminating source (twilight sky
or dome lamps) and the color of the night sky
across our extremely wide unfiltered band. Good superflats were obtained
provided that two requirements were met -- a lack of variable
illumination (e.g., by the Moon), and inclusion of
some minimum number of different fields in the superflat stack.
We found that at least eight different fields were needed to
create an acceptable superflat calibration frame. Moon illumination was
generally not a problem since most of the observations were obtained
close to new moon. While reducing data obtained during nights 
in which a sizable fraction of the images
were moderately affected by moonlight, we have found that using two
superflat frames (constructed from dark and slightly illuminated images
separately) provided a good solution for most of the data. 
A small number of images which were strongly affected by
moonlight were useless for our search and were discarded. 


Debiased and flattened images were registered to a reference frame defined by
previous images of each field using a triangulation algorithm developed by Giveon
(2000)\footnote{available from ftp://wise-ftp.tau.ac.il/pub/eran/IRAF/xyshift.f .}.
When needed, additional
correction for field rotation (due to inaccurate setting of the telescope rotator
angle) was performed using the {\tt geomap} and {\tt geotran} tasks within IRAF
\footnote{IRAF (Image Reduction and Analysis Facility) is distributed
by the National Optical Astronomy Observatories, which are operated
by AURA, Inc., under cooperative agreement with the
National Science Foundation.}.
Finally, individual sub-exposures of each field were summed
to obtain the final, $600$ s frame. 

\subsubsection{Image Subtraction} 

In order to discover variable or transient objects, we compared every new image
with a previously obtained reference frame. To initiate efficient follow-up
observations of interesting candidate objects (both imaging and spectroscopy;
see below), we needed to handle large amounts of data (several tens of fields
surveyed each night of the project) and to detect variable sources during each
observing run or shortly thereafter. This required the construction of an
automated analysis pipeline.  
We used image subtraction to
detect variable sources in the images, with the
advantage that any constant, underlying light,
even if bright compared to the variable source, is removed.
The required matching of the point-spread functions (PSFs)
of new and reference images was done using the
IRAF task {\tt psfmatch}.
As this method includes a division in Fourier space, it introduces
noise at high spatial frequencies that are poorly sampled by the
discrete data. We found that, in the case of WOOTS images, this noise
was manifested
as residuals in the centers of bright galaxies and stars, limiting our
ability to detect faint nuclear variability in galaxies and increasing the
number of false candidates. However, the overall
performance was acceptable, and allowed good detection of non-nuclear sources
and of brighter sources in galaxy cores.  

An alternative algorithm has been presented by Alard \& Lupton (1998) and
Alard (2000). This method (commonly referred to as ``ISIS'') does not perform
calculations in Fourier space.
A software package implementing the ISIS algorithm was made publicly available
by C. Alard. We obtained this package in 2000, and experimented with it
during the last phases of WOOTS. We found that ISIS generally
performs image subtractions superior to the results obtained with
{\tt psfmatch}. However, its mode of operation as a ``black box,'' with the
user having almost no influence on the selections made by the program, caused
it to fail occasionally (e.g., due to selection of image sections that are
contaminated by saturated stars).  

Motivated by the inadequate performance of {\tt psfmatch}, and in particular, the
limited sensitivity of WOOTS data to nuclear variability due to residual
subtraction artifacts, we have also developed an alternative subtraction
algorithm, the common PSF method (CPM; Gal-Yam et al. 2004; see Appendix A).
Unfortunately, both of the new algorithms (CPM and ISIS) were not available
during most of the time WOOTS was in operation.       

Following the use of {\tt psfmatch} in our analysis pipeline, the
images, now having similar PSFs, were next linearly scaled in flux
using the IRAF task {\tt linmatch}. The reference image was then subtracted
from the new image to create a difference image. Variable or transient sources
appeared as positive or negative residuals in this image.
Image subtraction involves a $\sqrt 2$ increase in Poisson noise, in addition
to possible subtraction residuals. Therefore, the faintest variable sources
we could typically detect had variable flux equivalent to $R \approx 21.5$ mag.    

\subsubsection{Variability Detection} 

The difference images produced by the image subtraction pipeline were
systematically searched for residuals using the {\tt findstar} package
written by E.
Almoznino\footnote{See ftp://wise-ftp.tau.ac.il/pub/nan/findst.ps.gz .},
which proved superior to the more commonly used {\tt daofind}
IRAF task. Chip defects and residuals from saturated,
bright, or known variable stars were automatically excluded. For residuals
detected at the $3\sigma$ level in the difference images, we required that
a $1\sigma$ object be detected in all three sub-exposures that were combined
to create the new image. This cut rejected
the vast majority of cosmic-ray hits, as well as many of the faster-moving
Solar System objects, which appear at a different position in each sub-exposure.
The final candidate lists (between zero and a few
tens per field) were scanned by eye, and obvious artifacts rejected. The
remaining objects were further studied (e.g., by searching the same locations
in previous images) and all convincing candidates were flagged for follow-up
observations. 

\subsubsection{Follow-up Data} 

Our follow-up observations consisted of photometric confirmation using
the Wise Observatory $1~$m telescope, and spectroscopy with various
larger telescopes.  
Follow-up photometry of candidate variable or transient objects
was secured as part of the WOOTS program. As soon as we identified a good
candidate source (many times while the WOOTS observing run was
still in progress), we imaged the field again, in order to confirm
(or refute) the validity of the source. A field in which an interesting
object (usually a SN) was discovered was given a high priority for
imaging in future WOOTS runs. Thus, if an object was still visible
during subsequent runs (typically spaced by one month), its light curve
would be sampled further. Sources discovered
after a WOOTS run was already finished were often confirmed
on subsequent nights, courtesy of our colleagues at Wise.
All of our sources ended up having two or more photometry
points within $\sim$1 week of discovery.

For each source, we tried to get
at least one filtered observation to supplement our unfiltered
data and to enable us to report the magnitude of the object in a standard
system. Such $R$-band magnitudes (calibrated relative to nearby sources
from the USNO A.1 catalog) are reported in Table~\ref{sntable}. Unfortunately,
our data were usually not of sufficient quality to produce
meaningful light curves for our SNe (see Appendix A for the exception),
and typically consist of two
detections within $\sim1$ week, at similar flux levels, followed
by non-detections in later runs. 

Prompt spectroscopy was used to identify the SN types (see below),      
utilizing a variety of telescopes, as listed in Table~\ref{spectable}. For
a few events, late-time spectroscopy (obtained long after the SN had
faded) was obtained using the low-resolution imaging spectrometer
(LRIS; Oke et al. 1995) mounted on the Keck-I 10~m telescope, in order
to determine the host-galaxy redshift or to identify AGN activity.  

\section{Results} 

\subsection{Supernovae} 

\subsubsection{SN Identification} 

During the course of WOOTS, we discovered many variable and transient
objects. The selection
of a clean sample containing only SNe was performed in the following manner. First,
we rejected variable point sources (i.e., sources that appear point-like in all
available epochs, and vary) which are either variable stars or AGNs. We were
left with candidates that appeared either as new point sources superposed on or near
galaxies, or transient sources that appeared for a short period of time and then
faded away (with no apparent host galaxy). Both classes of sources contain a
mixture of SNe and AGNs.  

The first class contains ``regular'' SNe superposed on their host
galaxies, and AGNs with resolved hosts. The second class (which is
less common) contains ``hostless'' SNe, either cluster intergalactic
SNe (2 such events detected; Gal-Yam et al. 2003a) or possibly SNe
whose hosts are too faint to detect in our imaging data (see, e.g.,
Strolger et al. 2002; the WOOTS sample, as it turns out, did not
contain such events). This class also contains AGNs below our
detection limit that become visible during bright flares. Such a
faint, variable, background AGN was projected near a foreground galaxy
in the field of the cluster Abell 24 and was initially reported as SN
1999de.  The AGN nature of the event became evident from follow-up
imaging obtained with the Keck I telescope in 2005, and confirmed
spectroscopically (Gal-Yam 2005). 

We rejected some AGNs from the above sample using their photometric
properties. If follow-up imaging showed that a nuclear SN candidate
behaves in a manner which is not expected of a SN (i.e., it exhibits an
extended period of optical variability containing several
rebrightenings), it was flagged as an AGN. Follow-up spectroscopy of
several such examples always showed that this was indeed the case. We did
not reject any SN candidate, though, based on its nuclear location alone. 
Final distinction between SNe and AGNs required spectroscopic
follow-up observations. These were preferentially obtained
shortly after discovery, and allowed us to confirm the SN nature of candidates
and to determine the SN type and redshift. Detailed discussion of
SNe 1998fc and 2001al is presented by Gal-Yam et al. (2003a), and their spectra
are shown there (Figures 3.1 and 3.2). An initial determination of
the SN redshifts, ages, and types (Gal-Yam 2003) was done by comparing the
observed spectra to redshifted versions of high signal-to-noise ratio (S/N)
template spectra of nearby SNe, of various types and ages (see Filippenko
1997 for a review),
drawn from the spectroscopic archive presented by Poznanski et al. (2002).
A detailed description of the procedure used is given by Gal-Yam et al. (2003a).
Usually, a second- or third-order polynomial was fitted and subtracted from both
the spectra and the templates,
in order to remove continuum variations due to instrumental and reddening effects,
as well as (crudely) any host-galaxy contamination.
 
For the final analysis we present here, we have employed the
{\tt Superfit} SN spectral identification software
package (Howell et al. 2005) in the following manner.
When possible, we have determined the host-galaxy redshift from narrow emission
lines. Next, we have run {\tt Superfit} on each spectrum using the measured
redshift (when known) or a large range of redshifts in all other cases. The
software was set to investigate all SN types and to attempt to fit an unknown
extinction ($A_V<2$ mag) and host-galaxy contamination. The best-fit results
are presented in Figures ~\ref{snspecfigfirst}--\ref{snspecfiglast} below.
In all but two cases, the identifications are obvious, and confirm the
analysis of Gal-Yam (2003).  

The spectra of SN 1999cg and SN 1999ct are noisy, and strongly
contaminated by host-galaxy light. Comparison of these spectra to that
of SN 1999ci, which combines the signal from a confirmed old SN~Ia
with significant contamination from a cluster galaxy, reveals strong
similarity, and suggests that these are also cluster SNe Ia (Gal-Yam
2003), but our {\tt Superfit} analysis does not provide conclusive
identifications. We have therefore obtained high S/N follow-up
spectroscopy of the host galaxies of these SNe
(Fig.~\ref{sn1999cghostspecfig}, Fig.~\ref{sn1999cthostspecfig}).  The
data confirm that these are early-type galaxies, lacking signatures of
recent star formation, at the cluster redshifts. Non-Type-Ia SNe are
rarely observed in such galaxies.  In view of their photometric
behavior and the lack of any AGN signatures, we thus conclude
that both SN 1999cg and SN 1999ct were most likely SNe~Ia. 



Spectroscopy of all other candidates, beyond the 12 SNe mentioned above,
showed that they were AGNs. Images of the
SNe are shown in Fig.~\ref{snfigmos1},
and their properties are summarized in Table~\ref{sntable}.
Spectroscopic observations are reported in Table~\ref{spectable} and
shown in Figures~\ref{snspecfigfirst}--\ref{snspecfiglast}.
A brief description of the AGNs we found is provided below (\S~\ref{agnsection}).  

\begin{table}
\caption{Supernovae Discovered by WOOTS}
\vspace{0.2cm}
\begin{centering}
\begin{minipage}{140mm}
\begin{scriptsize}
\begin{tabular}{llclcccccc}
\hline
SN & Type  & Redshift  & Cluster    & RA \& Dec   & Offsets         & \multicolumn {3} {c} {Discovery}   & Reference$^g$  \\
   &       & $z$       & (Redshift) & (J2000)$^a$ & (from host)$^b$ & $R$ [mag]$^c$ & Date$^d$ & Template$^e$ & \\   
\hline
1998cg & Ia & 0.119 & A1514 (0.199) & $12^h18^m18^s.13$        & $0.''4$ E & 18.5 & 980501 & 980401 & (1) \\
       &    &       &               & $+20^{\circ}44'26''.1$   & $4.''0$ N &      & 980513 &        &      \\
1998eu & Ia & 0.181 & A125 (0.188)  & $00^h59^m58^s.66$        & \nodata   & 20.7 & 981114 & 980724 & (2) \\
       &    &       &               & $+14^{\circ}18'00''.4$   & \nodata   &      & 981117 &        &      \\
1998fc$^f$ & Ia & 0.1023& A403 (0.103)  & $02^h59^m12^s.61$    & \nodata   & 20.5 & 981220 & 980114 & (3) \\
       &    &       &               & $+03^{\circ}29'39''.0$   & \nodata   &      & 990107 &        &      \\
1998fd & Ia & 0.24  & Field         & $01^h19^m18^s.06$        & $5.''8$ E & 21.3 & 981224 & 980115 & (3) \\
       &    &       &               & $+15^{\circ}55'24''.1$   & $2.''8$ N &      & 990109 &        &      \\
1999C  & Ia & 0.125 & A914 (0.195)  & $10^h08^m51^s.21$        & $0.''9$ E & 19.8 & 990114 & 980331 & (4) \\
       &    &       &               & $+71^{\circ}10'42''.9$   & \nodata   &      & 990115 &        &      \\
1999ax & Ia & 0.05  & A1852 (0.181) & $14^h03^m57^s.90$    & $2.''9$ W & 18.5 & 990320 & 980625 & (5) \\
       &    &       &               & $+15^{\circ}51'09''.1$   & $2.''3$ N &      & 990407 &        &      \\
1999ay & IIb & 0.044 & A1966 (0.151)& $14^h44^m43^s.86$        & $2.''8$ W & 18.0 & 990321 & 980727 & (5) \\
       &    &       &               & $+58^{\circ}55'42''.3$   & $2.''1$ S &      & 990322 &        &      \\
1999cg & Ia & 0.135 & A1607 (0.136) & $12^h43^m17^s.55$        & \nodata   & 20.0 & 990415 & 990320 & (6) \\
       &    &       &               & $+76^{\circ}13'58''.9$   &  \nodata  &      & 990512 &        &      \\
1999ch & Ia & 0.15  & A2235 (0.151) & $16^h54^m45^s.69$        & $0.''5$ W & 19.8 & 990513 & 990502 & (6) \\
       &    &       &               & $+39^{\circ}59'13''.9$   & \nodata   &      & 990516 &        &      \\
1999ci & Ia & 0.12  & A1984 (0.124) & $14^h52^m12^s.70$        & \nodata   & 20.4 & 990515 & 980527 & (6) \\
       &    &       &               & $+27^{\circ}54'22''.2$   &  \nodata  &      & 990523 &        &      \\
1999ct & Ia  & 0.180 & A1697 (0.183) & $13^h13^m04^s.56$        & $2.''8$ W & 21.2 & 990613 & 990215 & (7) \\
       &    &       &               & $+46^{\circ}15'52''.7$   & $1.''0$ N &      & 990615 &        &      \\
2001al$^f$ & Ia& 0.0723&A2122/4 (0.0661)& $15^h44^m51^s.74$    & \nodata   & 21.4 & 010326 & 000820 & (8) \\
       &    &       &               & $+36^{\circ}07'29''.0$   & \nodata   &      & 010328 &        &      \\
\hline
\end{tabular}
\end{scriptsize} 
Notes:\\
$^a$ Astrometry was performed with respect to nearby USNO-A sources, with
typical precision better than $0\farcs3$ in each coordinate.\\
$^b$ Blank entry indicates SN superposed on the nucleus of a
compact host for which no offsets are detected.\\
$^c$ Photometry was calibrated onto an $R$-band grid of nearby USNO-A sources, with
an average scatter of $0.2$ mag. However, individual values may suffer
up to $\sim$0.5 mag offsets due to uncertainties in the USNO-A
plate-to-plate calibration (Ofek 2000).\\
$^d$ Dates of discovery (upper) and confirmation (lower).\\
$^e$ Date of the latest image (obtained prior to the discovery
date) in which the SN does not appear.\\
$^f$ No host galaxy detected. \\
$^g$ References: 1. IAUC 6917 (Gal-Yam \& Maoz 1998a); 2. IAUC 7055 (Gal-Yam \& Maoz 1998b); 3. IAUC 7082 (Gal-Yam \& Maoz 1998a);
4. IAUC 7088 (Gal-Yam \& Maoz 1999b); 5. IAUC 7130 (Gal-Yam et al. 1999); 6. IAUC 7181 (Gal-Yam \& Maoz 1999c);
7. IAUC 7210 (Gal-Yam \& Maoz 1999d); 8. IAUC 7607 (Gal-Yam \& Maoz 2001).\\
\end{minipage}
\end{centering}
\label{sntable}
\end{table} 

\begin{table}
\caption {Supernova Spectroscopy} \label {Spec table}
\vspace{0.2cm}
\begin{centering}
\begin{minipage}{140mm}
\begin{scriptsize}
\begin{tabular}{lllccc}
\hline
Observed & Telescope + &  Observers & Observing   & Estimated  &  Reference$^b$  \\
SN       & Instrument  &            & Date        & SN Age$^a$ &  \\   
\hline
1998cg & ESO 3.6m + EFOSC2 & Patat \& Turatto              & 980530     & $\sim$33 & (1) \\
1998eu & AAT 4m + RGO      & Stathakis, Cannon, \& James   & 982111     & $\sim$24 & (2) \\
1998fc & ESO 3.6m + EFOSC2 & Leisy, Hainaut, \& Sekiguchi  & 990114     & \nodata  & (3) \\
       & Keck II + LRIS    & Filippenko, Leonard, \& Riess & 990120     & 25--42    & (4) \\
1998fd & Keck II + LRIS    & Filippenko, Leonard, \& Riess & 990119     & $\sim$49 & (4) \\
1999C  & Keck II + LRIS    & Filippenko, Leonard, \& Riess & 990120     & $\sim$14 & (5) \\
1999ax & KPNO 4m + RC      & Aldering                      & 990407     & $\sim$28 & (6) \\
       & AAT 4m + RGO      & Stathakis                     & 990409     & \nodata  & (6) \\
1999ay & KPNO 4m + RC      & Aldering                      & 990407     & $\sim$23 & (6) \\
1999cg & Keck II + LRIS    & Sorenson \& Schaefer          & 990530     & \nodata  & (7) \\
(host) & Keck I + LRIS     & Gal-Yam \& Sharon             & 070716     & \nodata  & (8) \\
1999ch & KPNO 4m + RC      & Pogge                         & 990608     & $\sim$19 & (9) \\
1999ci & KPNO 4m + RC      & Pogge                         & 990609     & $\sim$22 & (9) \\
1999ct & Keck II + LRIS    & Guhathakurta                  & 990806     & \nodata  & (10) \\
(host) & Keck I + LRIS     & Gal-Yam \& Sharon             & 070716     & \nodata  & (8) \\
2001al & Keck I + LRIS     & Filippenko, Barth, \& Leonard & 010329     & 21--33    & (11) \\
\hline
\end{tabular}
\label{spectable}
\end{scriptsize} 
Notes:\\
$^a$ SN age, at the time of spectroscopy, given in days relative to $B$-band maximum.\\
$^b$ References: 1. IAUC 6925 (Patat \& Turatto 1998); 2. IAUC 7082 (Gal-Yam \& Maoz 1999a); 3. IAUC 7093 (Gal-Yam \& Maoz 1999e);
4. IAUC 7091 (Filippenko, Leonard, \& Riess 1999a); 5. IAUC 7092 (Filippenko, Leonard, \& Riess 1999b);
6. IAUC 7357 (Gal-Yam \& Maoz 2000c); 7. IAUC 7198 (Gal-Yam, Maoz, \& Guhathakurta 1999); 8. This work;
9. IAUC 7199 (Gal-Yam, Maoz, \& Pogge
1999); 10. IAUC 7356 (Gal-Yam, Maoz, \& Guhathakurta 2000);
11. IAUC 7607 (Filippenko, Barth, \& Leonard 2001).\\
\end{minipage}
\end{centering}
\end{table} 

\begin{figure*}
\includegraphics[width=100mm]{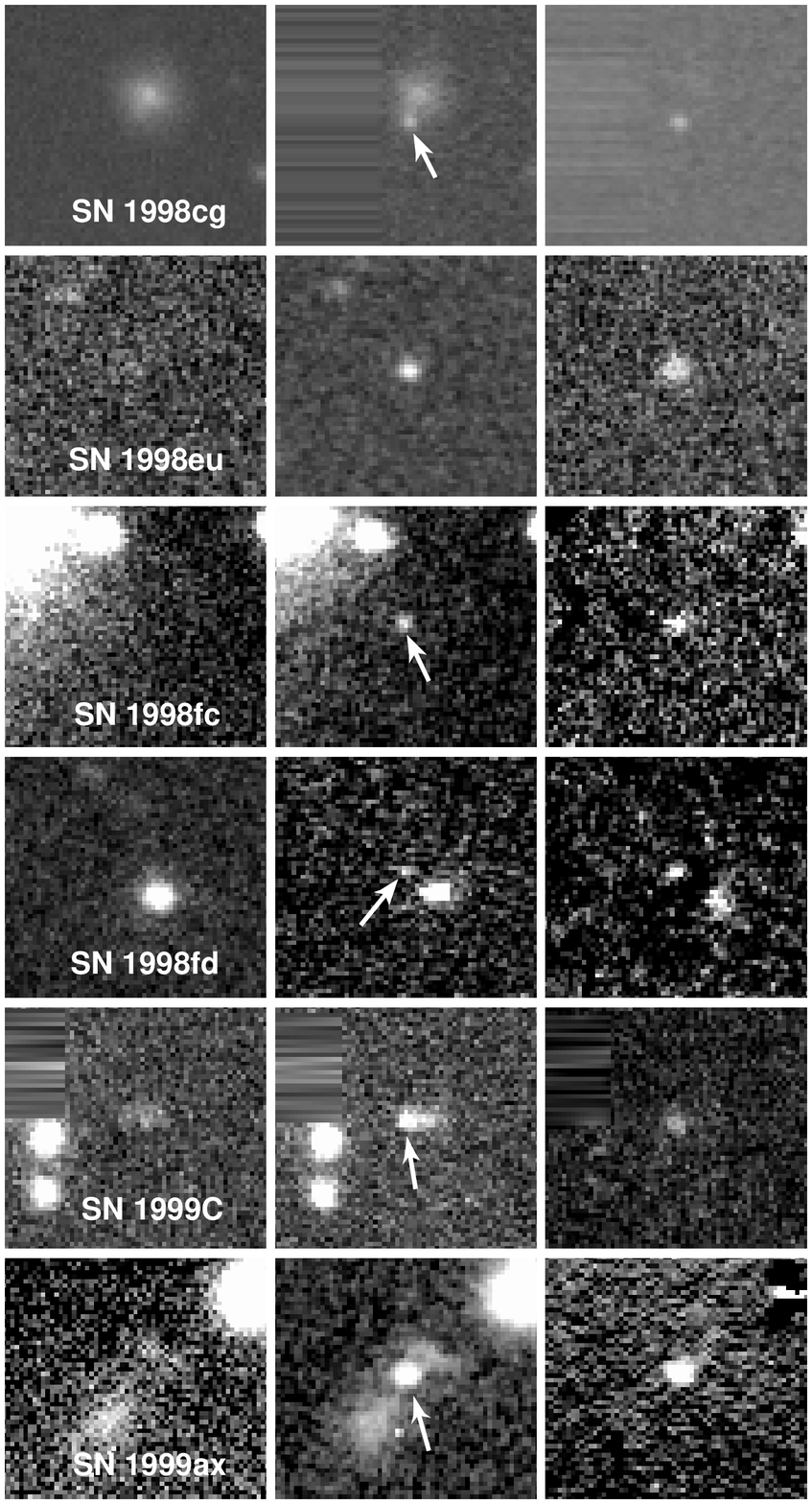}
\end{figure*}

\begin{figure*}
\includegraphics[width=100mm]{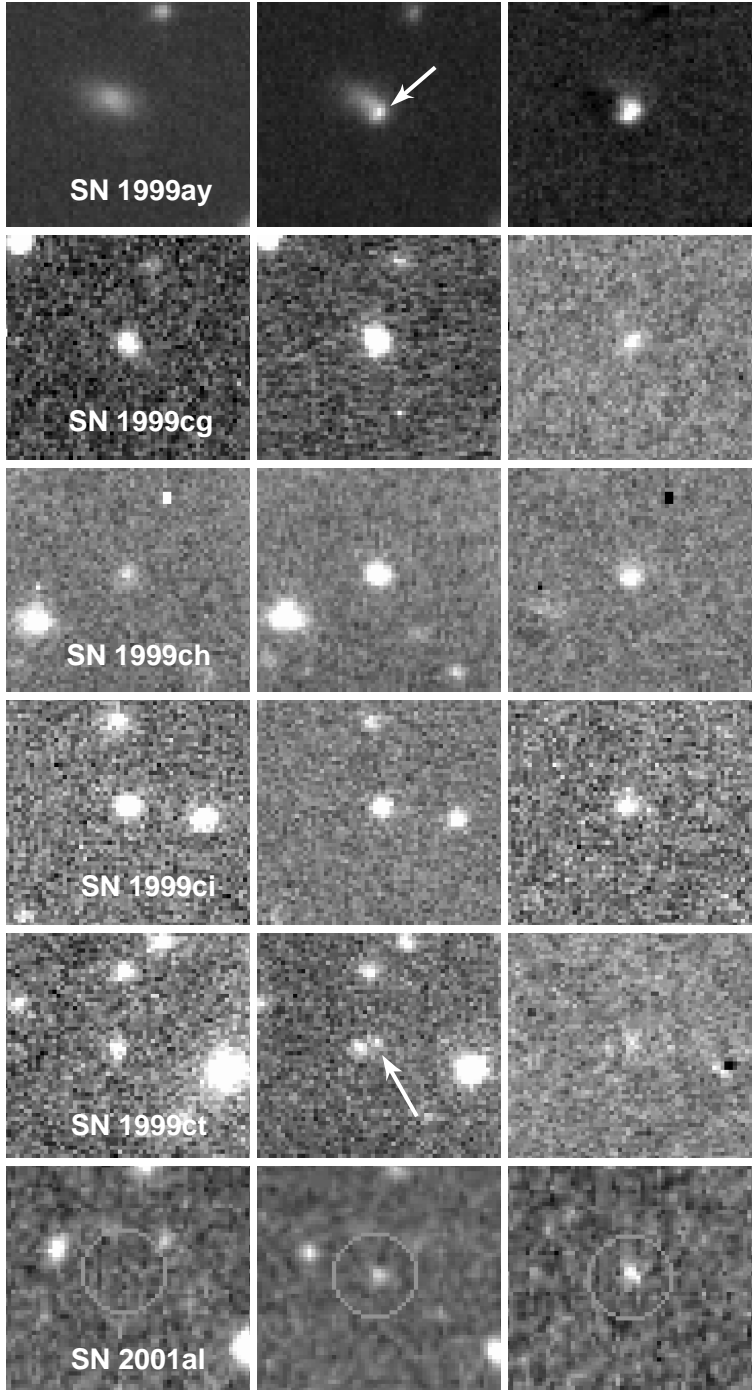}
\caption{WOOTS discovery images of supernovae. The right-hand panels show the
difference frames produced by subtracting a pre-explosion reference frame
(left-hand panels) from the SN discovery images (middle panels). Arrows
point out SNe that are well-resolved from their host-galaxy nuclei.
Each panel is $40''$ on a side; north is up and east to the left. Striped
regions are devoid of data, due to CCD edges or regions affected by bright
stars.}
\label{snfigmos1}
\end{figure*} 

\begin{figure}
\plotone{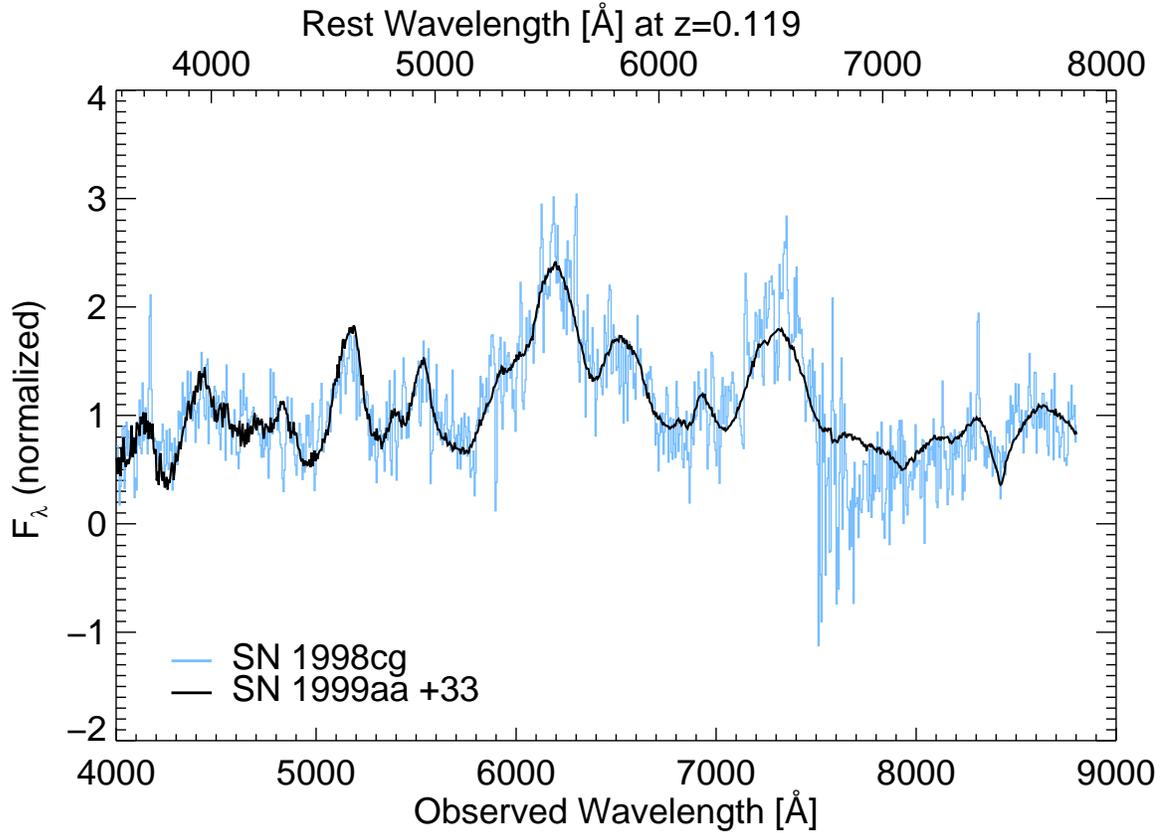}
\caption{A spectrum of SN 1998cg compared with a high S/N template SN~Ia
spectrum. Investigation using {\tt Superfit} indicates negligible galaxy
contamination, and a preference for template spectra of events belonging
to the slow/luminous (SN 1991T-like; Filippenko et al. 1992a) subclass. Here
and in Figures \ref{snspecfigfirst}--\ref{snspecfiglast}, the thin dark curve
is the best-fit spectrum found by {\tt Superfit} (with the SN name and age
relative to $B$-band maximum marked), and the grey curve is the observed
WOOTS spectrum.}
\label{snspecfigfirst}
\end{figure} 

\begin{figure}
\plotone{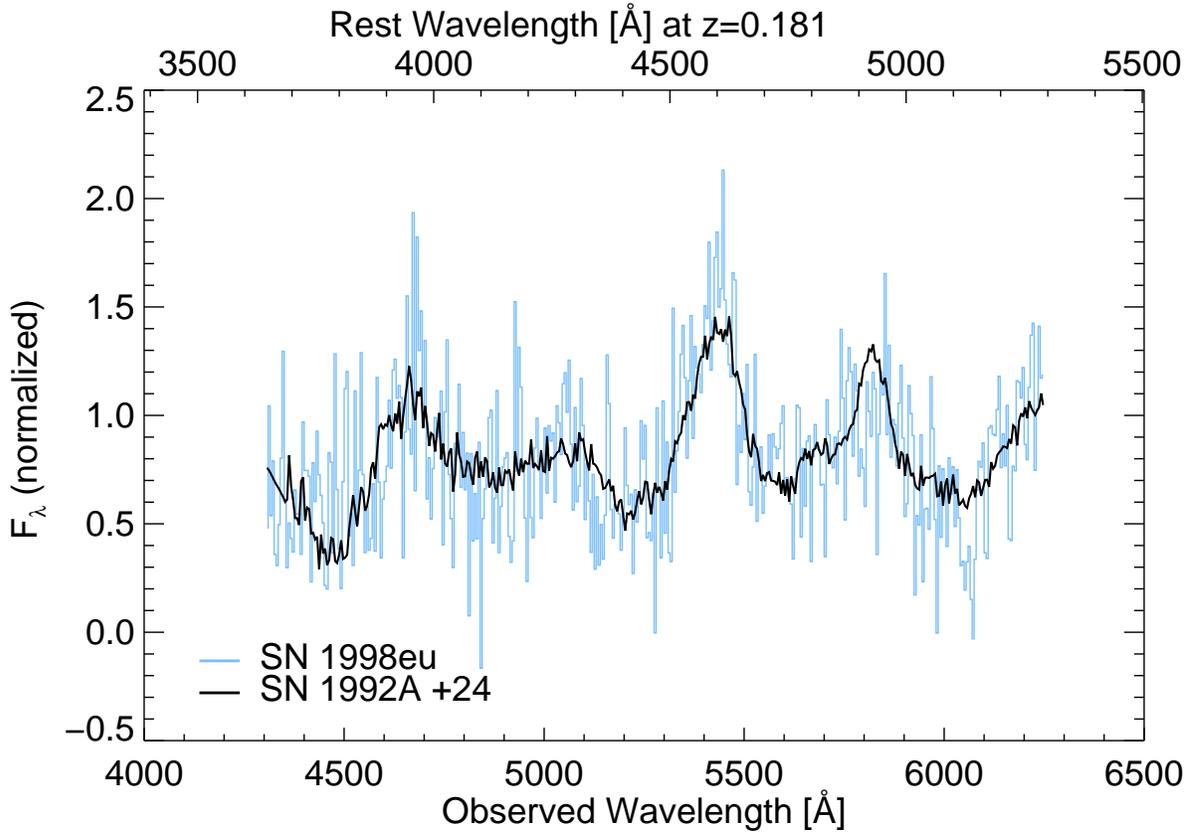}
\caption{A spectrum of SN 1998eu compared with a high S/N template.
Normal SNe Ia provide the best fit, and the {\tt Superfit} analysis
suggests a star-forming host galaxy with a spectral type (Sc) consistent
with the blue colors (Table~\ref{hosttable}) and strong emission
lines seen in the raw spectrum.}
\end{figure} 

\begin{figure}
\plotone{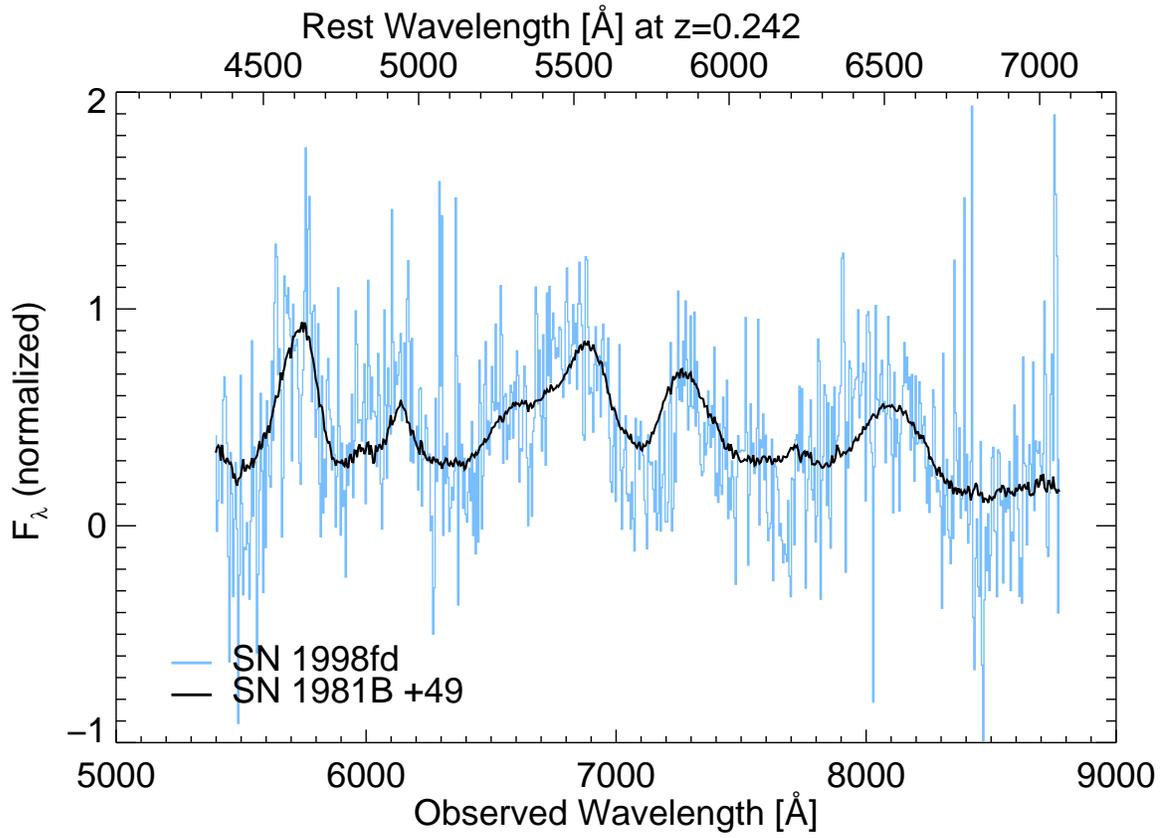}
\caption{A spectrum of SN 1998fd compared to a high S/N template.
Normal SNe Ia provide the best fit. Host-galaxy properties are not
significantly constrained.}
\end{figure} 

\begin{figure}
\plotone{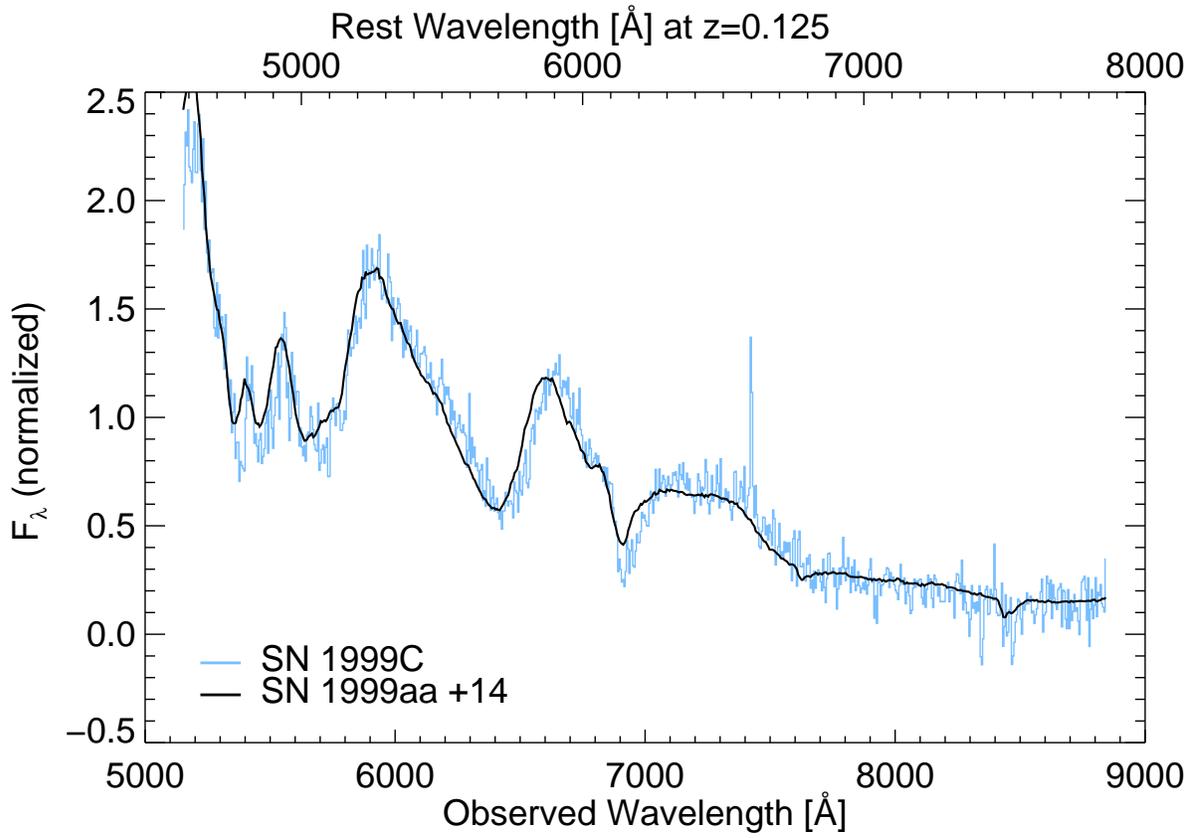}
\caption{A spectrum of SN 1999C compared with a high S/N template.
Luminous (SN 1991T-like) SNe Ia provide the best fit. The preferred
host-galaxy spectrum by {\tt Superfit} suggests an early-type spiral,
consistent with the detection of weak H$\alpha$ but no oxygen
emission lines in the host spectrum.}
\end{figure} 

\begin{figure}
\plotone{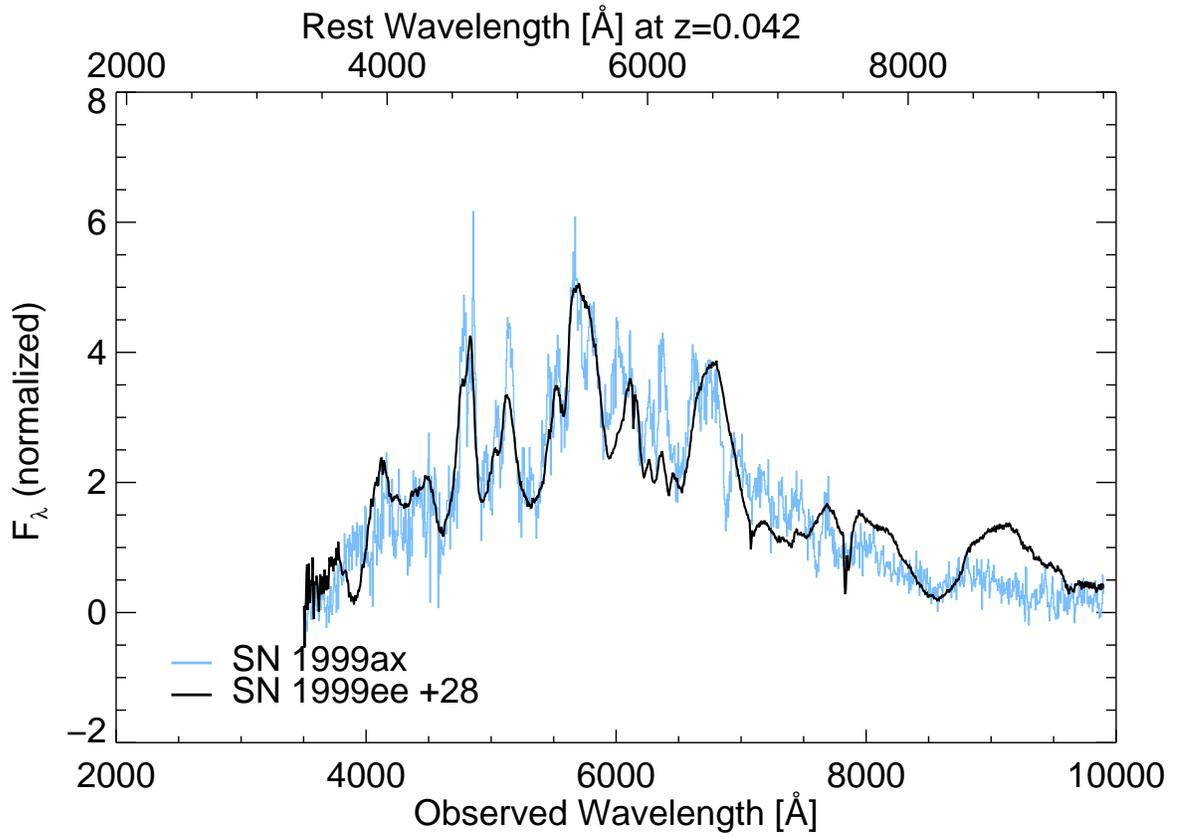}
\caption{A spectrum of SN 1999ax compared to a high S/N template.
Normal SNe Ia provide the best fit, but we note that the spectrum is
somewhat peculiar, showing a weak Ca near-IR triplet
and odd structure around 6000~\AA. Host-galaxy contamination is negligible.}
\end{figure} 

\begin{figure}
\plotone{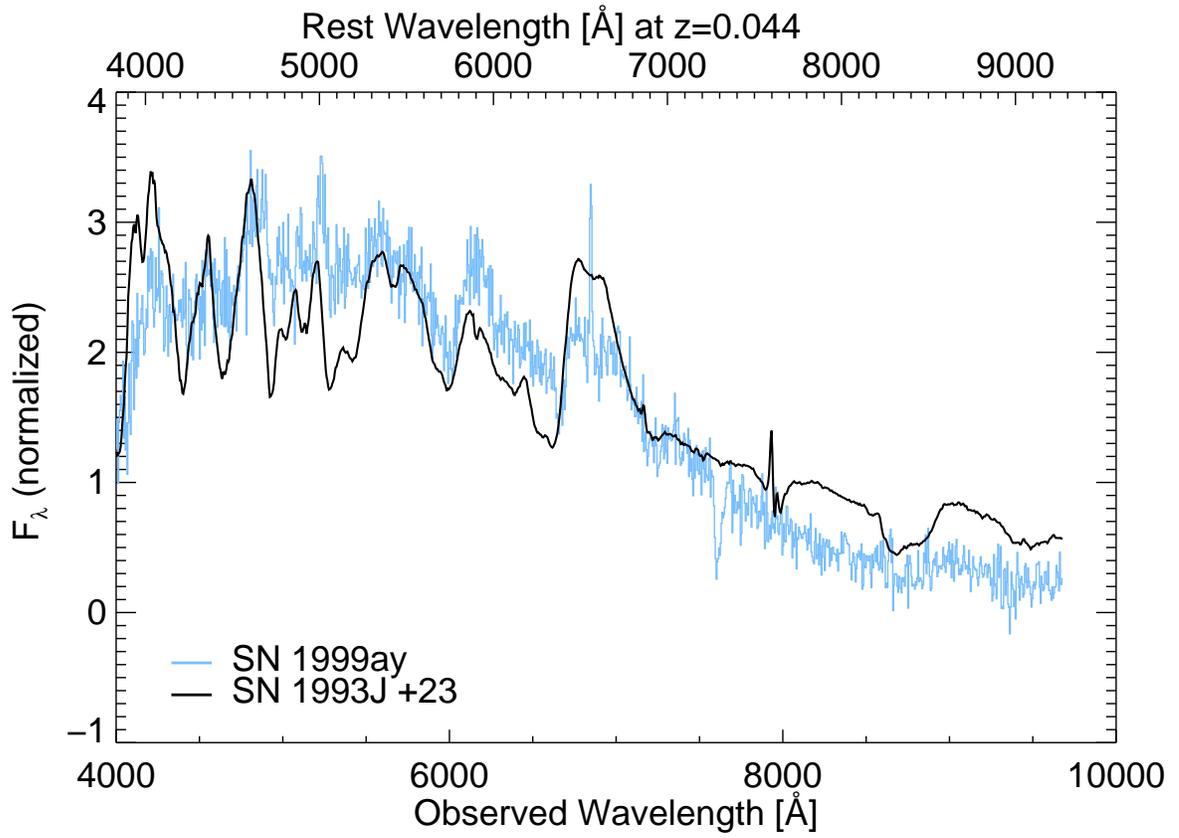}
\caption{A spectrum of SN 1999ay compared to a high S/N template.
This appears to be a SN~IIb similar to the prototype SN 1993J (as shown).
The SN photometry (Appendix A) indeed shows the expected rapid decline.}
\label{sn1999ay-spec-fig}
\end{figure} 

\begin{figure}
\plotone{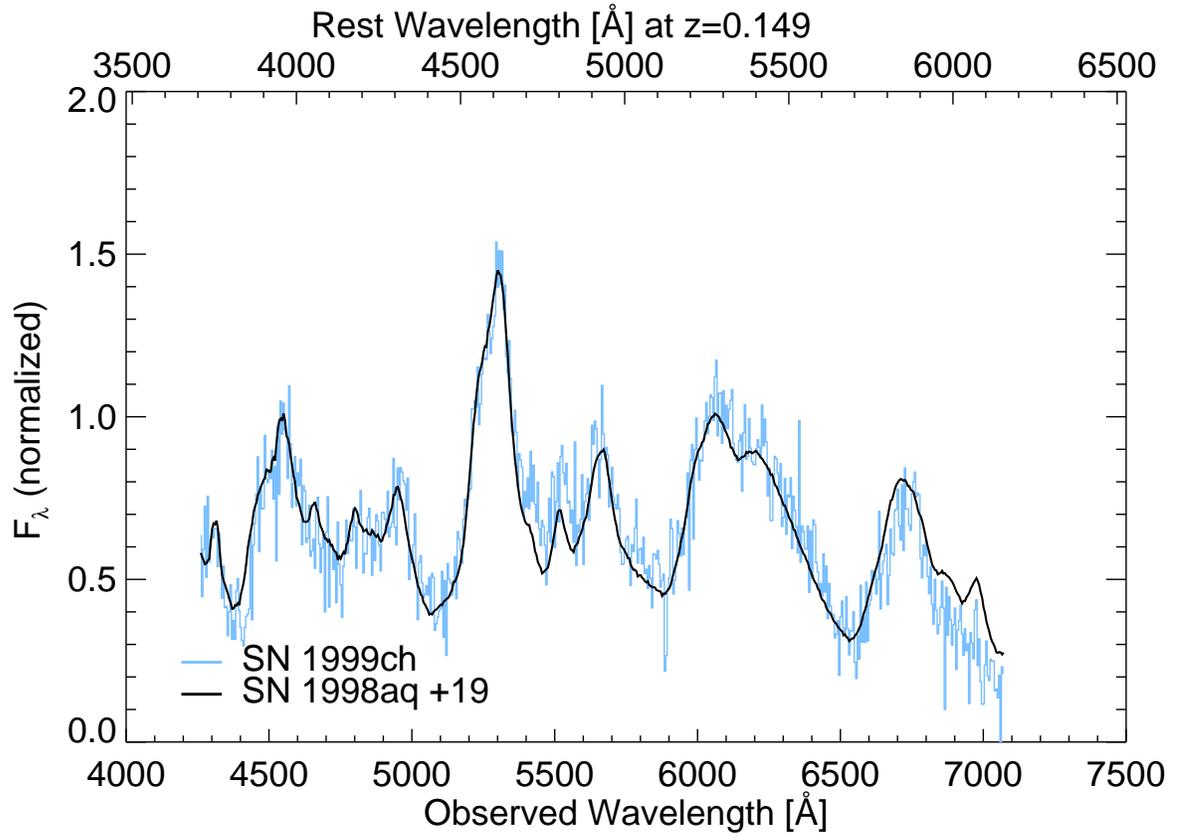}
\caption{A spectrum of SN 1999ch compared to a high S/N template.
{\tt Superfit} analysis suggests a normal SN Ia, while the host is best fit by
an early-type galaxy (S0/E).}
\end{figure} 

\begin{figure}
\plotone{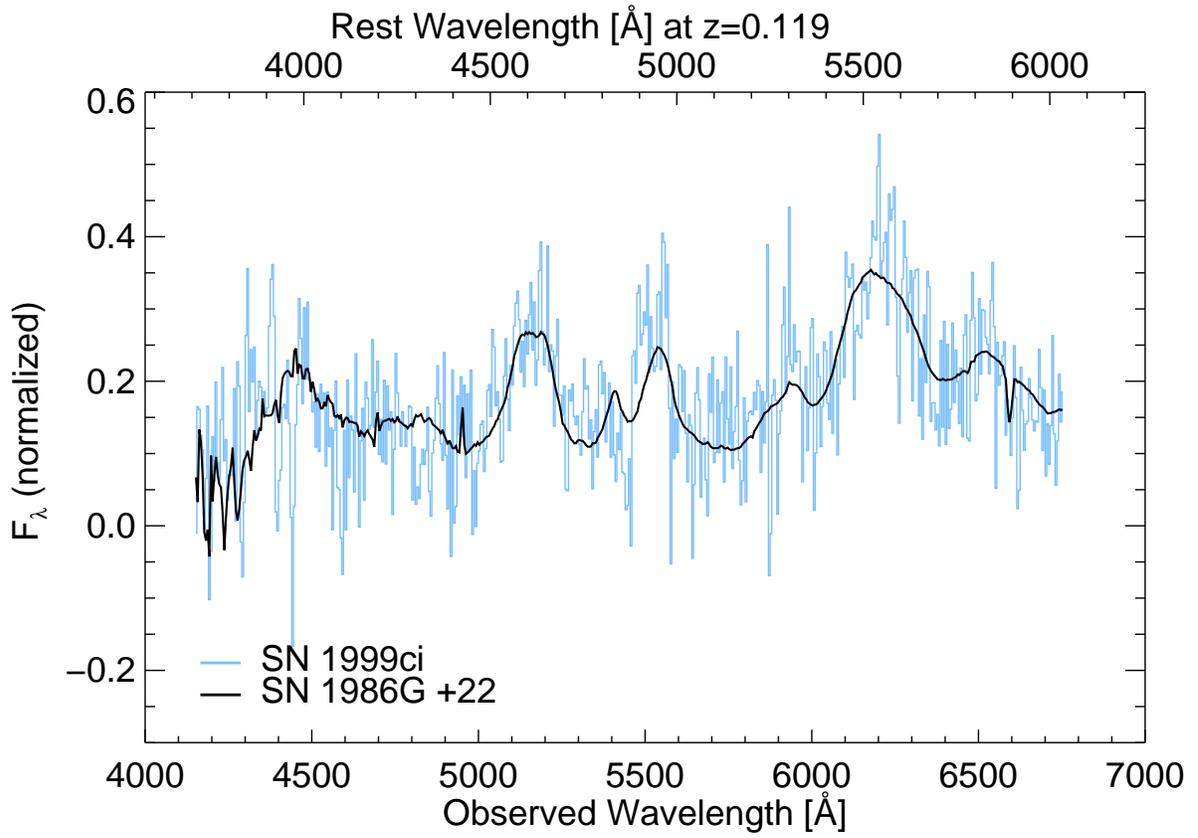}
\caption{A spectrum of SN 1999ci compared to a high S/N template.
{\tt Superfit} analysis suggests a normal or subluminous (SN
1991bg-like; Filippenko et al. 1992b) SN~Ia, heavily contaminated by
an early-type spiral host.}
\label{snspecfiglast}
\end{figure} 

\begin{figure}
\plotone{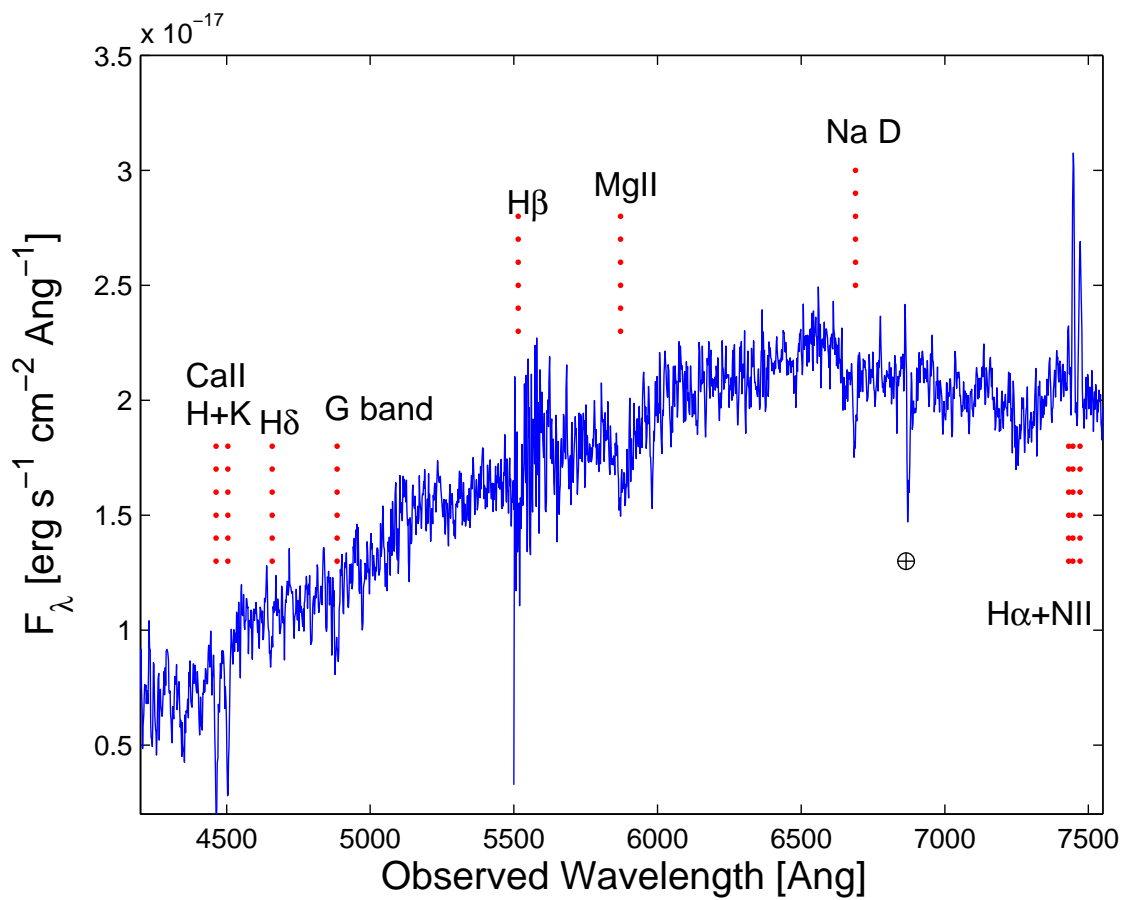}
\caption{A spectrum of the host galaxy of SN 1999cg. The spectral shape and
lack of prominent emission lines (H$\alpha$ is in emission, while all other Balmer
lines are in absorption) indicate little ongoing star formation, and an
early spectral class (Sa). The measured redshift based on numerous features
(marked) is $z=0.1345$, in excellent agreement with that of the cluster Abell 1607
($z=0.136$, Table 1).}
\label{sn1999cghostspecfig}
\end{figure} 

\begin{figure}
\plotone{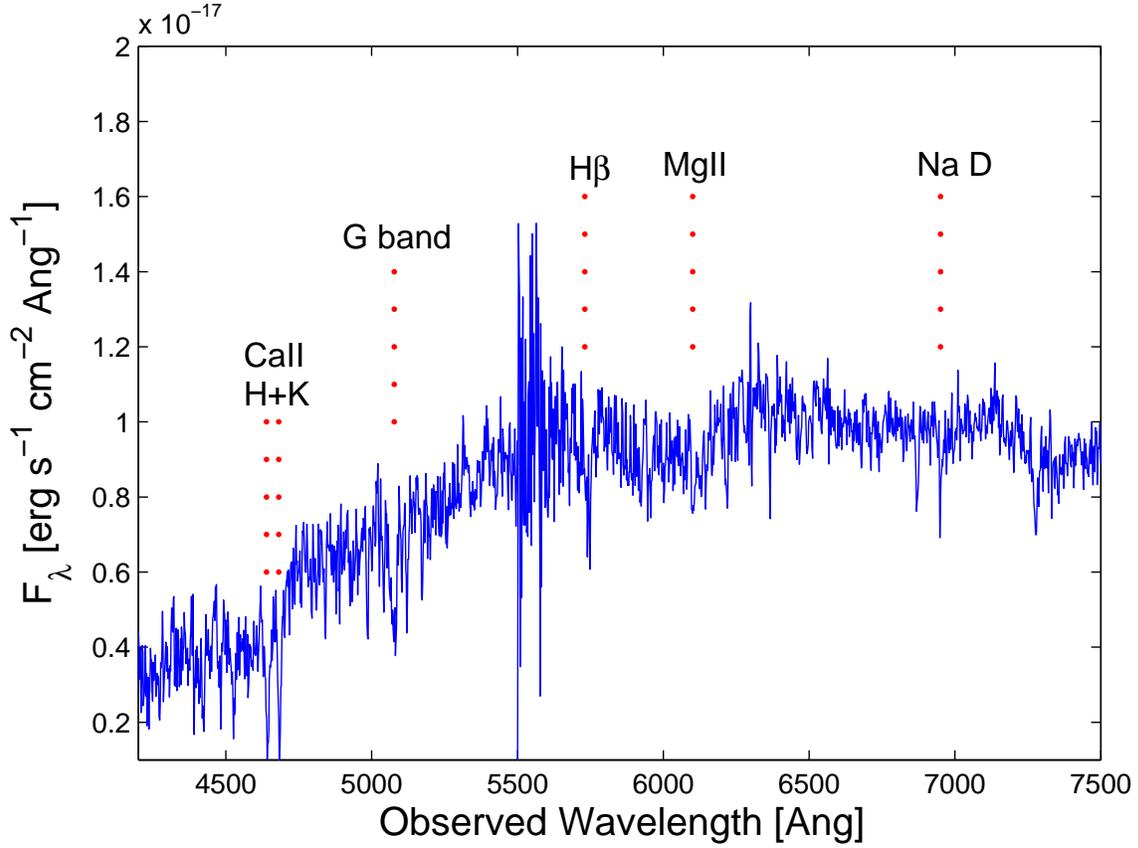}
\caption{A spectrum of the host galaxy of SN 1999ct. The spectral shape and
lack of any emission lines indicate little (if any) ongoing star formation, and an
early spectral class (E/S0). The measured redshift based on numerous features
(marked) is $z=0.180$, in good agreement with that of the cluster Abell 1697
($z=0.183$, Table 1).}
\label{sn1999cthostspecfig}
\end{figure} 

\subsubsection{The WOOTS SN Sample} 

In total, we discovered 12 SNe in the course of WOOTS. Of those,
seven occurred in cluster galaxies, while the rest were field events.
Eleven of the SNe, including all cluster events, are SNe~Ia, based on spectroscopy
and host-galaxy properties. Contamination by core-collapse SNe, even of SNe~Ic
that sometimes closely resemble SNe Ia, is unlikely in the cluster sample (due to the
early-type nature of the hosts) and also for the 4 field SNe Ia (which happen to have
relatively high S/N spectra). We identify the only non-SN~Ia event, SN 1999ay,
as a possible SN~IIb (Filippenko 1988; Filippenko, Matheson, \& Ho 1993)
at $z=0.04$--0.05 based on two spectra (Gal-Yam \& Maoz 2000c, and this work).
Its photometric behavior (Appendix A), and a redshift of 0.044 recently
measured by the Sloan Digital Sky Survey (SDSS; Adelman-McCarthy et al. 2005),
appear to support our identification.

For two of the cluster SNe we detected no visible hosts in WOOTS data and in
deep Keck images; these turned out to be the first demonstrable
cases of intergalactic SNe, which are discussed in detail by Gal-Yam et
al. (2003a). We found no background SNe for which there
was evidence of lensing by the clusters.  

WOOTS is the only survey we are aware of that produced a sample of SN candidates
which were all spectroscopically confirmed (or ruled out). Based on our statistics
of non-cluster SNe, in a flux-limited SN survey with a limiting magnitude
comparable to that of WOOTS (i.e., $\sim21.5$ mag in the $R$ band), 4/5 of
SNe discovered would be SNe~Ia. This estimate is of course crude,
as it is based on so few events. However, it is supported by results
from other similar surveys. In particular, we find that SNe~Ia account for
$70\%$ (21/30) of the SNe discovered by the Lick Observatory Supernova
Search (LOSS) with the Katzman Automatic Imaging Telescope (Filippenko
et al. 2001; Filippenko 2005a; Filippenko \& Li 2008) in random galaxies
projected around LOSS
target galaxies. Since the vast majority of the nearby galaxies targeted
by LOSS cover only a small fraction of the KAIT field of view,
this program provides, in addition to the main SN survey in nearby galaxies, also
a blank-field survey covering small patches of surrounding sky to a limiting
magnitude of $R \approx 19.5$ mag, which is almost spectroscopically
complete (see Appendix B for more details). Results from the SN Factory
project (Weaver et al. 2006) lead to very similar estimates (P. Nugent,
2007, private communication).  

Comparing the above results with the reported fraction of SNe~Ia from
the total number of field events in the Mount Stromlo Abell Cluster
SN Search (MSACSS; $<9/21$; Table 1 of Germany et al. 2004) shows a
puzzling discrepancy. A likely explanation is that SN classification
in MSACSS, which relied heavily on light-curve fitting due to the
low fraction of SNe with spectroscopic observations, was compromised.
Perhaps a significant fraction of the events that Germany et al. (2004)
classify as ``nIa'' (non-SNe~Ia, based on light-curve shape) are
in fact SNe~Ia, which either have peculiar light curves, or poorer
photometry than the authors realized, or are not SNe at all (e.g.,
these are AGNs).

The ratio of SNe Ia to core-collapse (non-Ia) events reflects the
strong bias favoring the discovery of luminous SNe~Ia, rather than the
relative rates (per unit volume) of SNe. We also note that for a
program with similar properties (depth and field size), a cluster
survey increases the number of detected SNe by a factor of $\sim2$
compared to blank-field imaging (similar to the results reported by
Reiss et al. 1998 and Germany et al. 2004, see below).  

Since the completion of WOOTS, data from the SDSS have become
available for most of our SN fields. In Table~\ref{hosttable} we
summarize the properties of the host galaxies of eight
WOOTS SNe as measured by the SDSS. We note that, with the exception of
SN 1998cg, which resides in a large field spiral, most SNe (both cluster
and field events) occurred in relatively low-luminosity (0.1- 0.15$L_*$)
galaxies. While field SNe occurred in blue ($g-r<0.6$ mag) galaxies,
cluster SNe typically occurred in red ($g-r>0.8$ mag) galaxies,
where we adopt $g-r=0.6$ mag, the minimum in the bimodal color
distributions of Blanton et al. (2003a, their Fig. 7), as the division
between red and blue galaxies. Curiously, however, cluster SN~Ia 1998eu
occurred in a faint blue galaxy at the redshift of the cluster Abell 125.  

\begin{table}
\caption{SDSS Data for Host Galaxies of WOOTS Supernovae}
\vspace{0.2cm}
\begin{centering}
\begin{minipage}{140mm}
\begin{scriptsize}
\begin{tabular}{lllllcllll}
\hline
SN host & \multicolumn {2} {c} {RA \& Dec} & \multicolumn {5} {c} {Photometry (mag)} & Luminosity$^a$ & Comments  \\
        & \multicolumn {2} {c} {(J2000)} & $u-r$ & $g-r$ & $r$ & $r-i$ & $r-z$ & $[L_r/L_*]$\\   
\hline
1998cg-host & $12^h18^m18^s.12$ & $+20^{\circ}44'30''.3 $ & 1.82 & 0.58 & 17.28 & 0.31 & 0.47 & 1.23 & Face-on spiral (field)\\
1998eu-host & $00^h59^m58^s.66$ & $+14^{\circ}28'00''.4 $ & 0.81 & 0.34 & 21.06 & 0.14 & $-$0.25 & 0.10 & Faint blue galaxy (cluster)\\
1998ax-host & $14^h03^m57^s.64$ & $+15^{\circ}51'11''.9 $ & 1.02 & 0.47 & 19.05 & 0.89 & 0.10 & 0.04 & Galaxy pair -- NW (field)\\
1998ax-host & $14^h03^m58^s.28$ & $+15^{\circ}51'01''.3 $ & 1.33 & 0.35 & 17.73 & 0.11 & $-$0.32 & 0.13 & Galaxy pair -- SE (field)\\
1998ay-host & $14^h44^m44^s.22$ & $+58^{\circ}55'43''.5 $ & 1.75 & 0.46 & 17.50 & 0.32 & 0.49 & 0.12 & Blue spiral (field)\\
1999ch-host & $16^h54^m45^s.71$ & $+39^{\circ}59'13''.9 $ & 2.72 & 0.82 & 20.37 & 0.35 & 0.70 & 0.12 & Faint compact galaxy (cluster)\\
1999ci-host & $14^h52^m12^s.60$ & $+27^{\circ}54'22''.7 $ & 2.56 & 0.80 & 19.62 & 0.41 & 0.74 & 0.16 & Faint early type (cluster)\\
1999ct-host & $13^h13^m04^s.83$ & $+46^{\circ}15'51''.7 $ & 1.78 & 0.98 & 20.68 & 0.51 & 0.82 & 0.14 & Faint early type (cluster)\\
\hline
\end{tabular}
\end{scriptsize} 
Notes:\\
$^a$ Luminosities are calculated assuming $L_{*}=-21.21$ mag for the SDSS $r$ band
from Blanton et al. (2003b), and the standard flat WMAP cosmology ($H_{0}=71$
km s$^{-1}$ Mpc$^{-1}$, $\Omega_{\Lambda}=0.73$).
Galactic extinction correction based on maps by Schlegel, Finkbeiner, \& Davis (1998) has been applied.\\
\end{minipage}
\end{centering}
\label{hosttable}
\end{table} 

\subsection{Other Transients}
\label{agnsection} 

\begin{figure}
\plotone{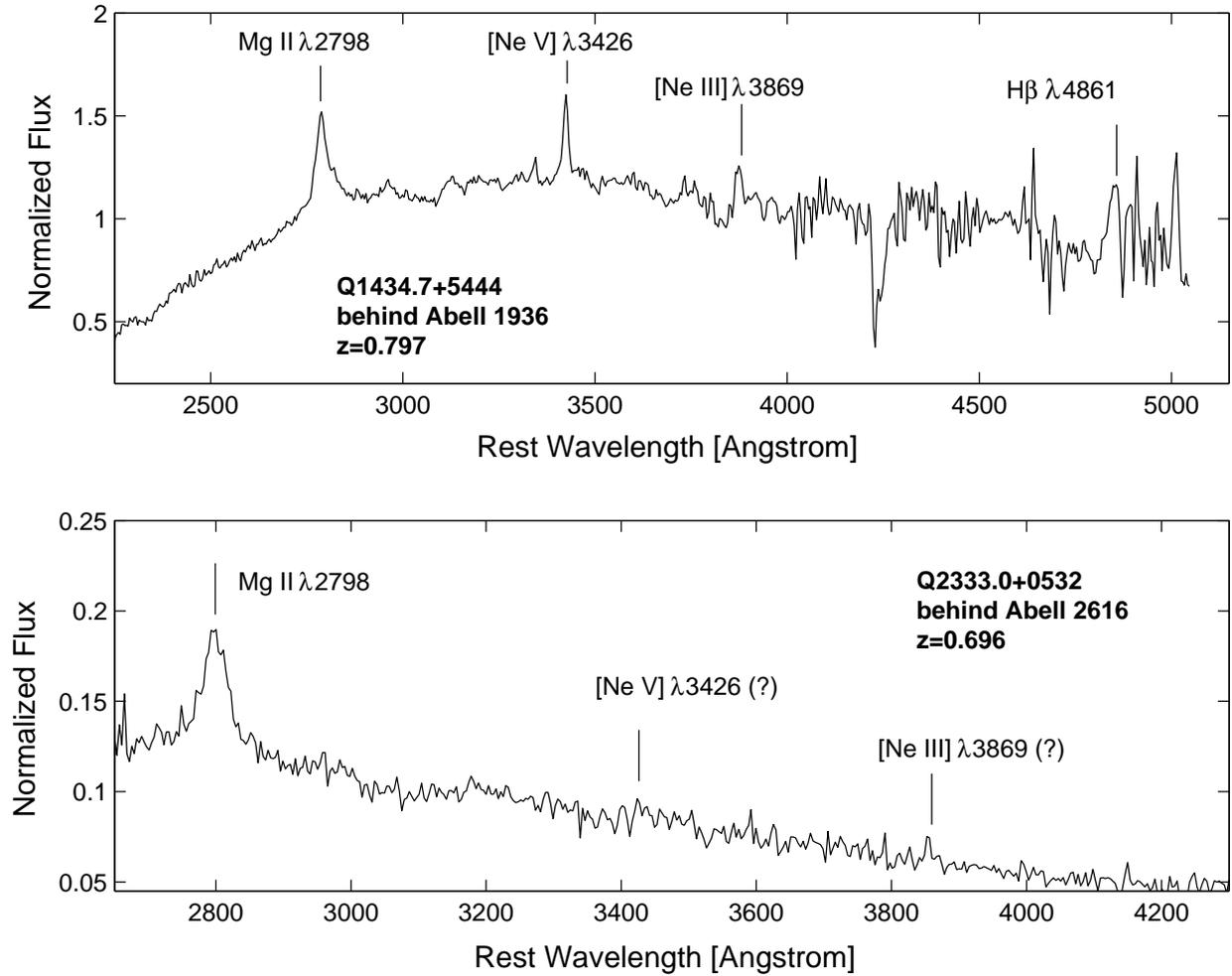}
\caption{Example Keck spectra of WOOTS AGNs behind galaxy clusters.}
\label{agnfig}
\end{figure}

A number of variable sources that are not SNe (based on their variability
properties) were also discovered in WOOTS data. Being lower-priority
targets, many of these sources were not followed up.
Some of them are Galactic variable stars of various
kinds, while others have turned out to be AGNs behind the target
clusters. Additional AGNs were spectroscopically culled from the SN sample
(see above). Example AGNs are listed in Table~\ref{agntable} and
example spectra are shown in Fig.~\ref{agnfig}. It is interesting to note
that all the AGNs we discovered reside at relatively low redshifts
($z\leq1$). However, our selection process (which rejected variable point
sources prior to spectroscopy, but retained resolved or marginally resolved
AGNs) is probably responsible for this effect, since high-$z$ (and thus more
luminous) AGNs would dominate their host, and thus appear star-like.
Lower-luminosity AGNs (analogs of local Seyfert galaxies), on the other hand,
which we could detect only out to lower redshifts, would better mimic
SNe (and gain spectroscopic follow-up observations) due to higher
contamination by their resolved host-galaxy light. 

One of the original goals of WOOTS was to discover nuclear variability
in galaxies due to AGNs in the clusters or to tidal disruption of stars
by central supermassive black holes in quiescent galaxies. 
Our ability to conduct such a search was compromised by
imperfection in the image-subtraction methods we initially used,
resulting in many spurious residuals in galaxy cores.
None of the AGNs we discovered turned out to be cluster members.  

\begin{table}
\caption{WOOTS Active Galactic Nuclei}
\vspace{0.2cm}
\begin{centering}
\begin{minipage}{140mm}
\begin{scriptsize}
\begin{tabular}{lcllllllll}
\hline
AGN  & Redshift  & \multicolumn {2} {c} {RA \& Dec} & \multicolumn {5} {c} {SDSS Photometry (mag)} & Reference$^c$  \\
     &   $z$     & \multicolumn {2} {c} {(J2000)}   & $u$ & $g$ & $r$ & $i$ & $z$ & \\   
\hline
WOOTSJ143444+544427 & 0.797 & $14^h34^m43^s.88$ & $+54^{\circ}44'26''.5 $$^b$ & 20.18 & 19.95 & 19.77 & 19.66 & 19.51 & (1)\\
WOOTSJ233304+053224 & 0.696 & $23^h33^m03^s.64$ & $+05^{\circ}32'23''.5 $ &  &  &  &  &  & (2)\\
WOOTSJ144409+585332 & 0.791 & $14^h44^m08^s.80$ & $+58^{\circ}53'31''.6 $$^b$ & 19.45 & 19.12 & 18.99 & 19.07 & 18.92 & (3)\\
WOOTSJ002731+214247 & 1.064 & $00^h27^m31^s.06$ & $+21^{\circ}42'47''.5 $ &  &  &  &  &  & (4) \\
WOOTSJ002228+232356 & 0.650 & $00^h22^m28^s.20$ & $+23^{\circ}23'56''.2 $ &  &  &  &  &  & (4) \\
WOOTSJ131315+462001$^a$ & \nodata & $13^h13^m15^s.42$ & $+46^{\circ}20'00''.8 $$^b$ & 20.21 & 20.00 & 19.94 & 19.89 & 19.76 & (5)\\
\hline
\end{tabular}
\end{scriptsize} 
Notes:\\
$^a$ No spectrum obtained of this source. However, its strong variability and blue
SDSS colors suggest an AGN with $z<2$.\\
$^b$ SDSS astrometry is reported for these sources rather than our original USNO-based
locations, which are, however, always within $0.3''$ of the SDSS coordinates.\\
$^c$ References: 1. IAUC 7356 (Gal-Yam, Maoz, \& Guhathakurta 2000); 2. IAUC 7228 (Matheson, Modjaz, \& Filippenko 1999);
3. IAUC 8171 (Gal-Yam et al. 2003c); 4. ATEL 586 (Gal-Yam 2005); 5. IAUC 7210 (Gal-Yam \& Maoz 1999d).\\
\end{minipage}
\end{centering}
\label{agntable}
\end{table} 

\subsection{Asteroids} 

The automated WOOTS analysis pipeline also detected asteroids
in the data. Due to their proper motion, asteroids
usually appear to be ``smeared'' in each sub-exposure, and
show detectable motion between each of the three sub-exposures
we normally obtained.  
Upon discovery of an asteroid by our
program, amateur astronomer I. Manulis performed astrometry using the WOOTS
images and reported these positions to the
International Astronomical Union Minor Planet Center (MPC).
When possible, the positions were used to calculate a preliminary
motion vector, and follow-up observations were obtained in
order to determine the orbit of each object.
In other cases, the positions we measured were linked by the
MPC to other observations and used to determine the
orbits of these objects.  
Over 50 asteroids
were reported, more than $90\%$ of them previously
unknown (see Manulis 2002 for more details).
Three objects have so far been numbered by the MPC.
We have found that the relatively deep limiting
magnitudes obtained by WOOTS (asteroids as faint as $R=20$
mag were measured) made our data unique in sampling a
faint population of main-belt asteroids that was not accessible to
other systematic surveys for Solar System objects active at the time. 
We did not discover any comets, slowly moving (Kuiper belt or
Trans-Neptunian) objects, or very fast (near-Earth) bodies.

\section{Comparison with Other Surveys} 
Numerous surveys for SNe and other variable and transient objects
have been carried out over the last few years. In particular, the
cosmological utility of SNe~Ia has been a driver for many dedicated SN
surveys, from nearby galaxies (e.g., LOSS, Filippenko et al. 2001;
Filippenko 2005a; Filippenko \& Li 2008) out to high redshifts
(e.g., the SN legacy survey,
Astier et al. 2006; ESSENCE, Wood-Vasey et al. 2007) and very high
redshifts (e.g., GOODS, Riess et al. 2007; SDFSNS, Poznanski et al.
2007). Most of these programs do not target clusters of galaxies.
Here, we briefly compare our survey with other cluster SN search programs.
     
The Mount Stromlo Abell Cluster SN Search (MSACSS; Reiss et al. 1998) was
a survey of similar design to WOOTS, targeting lower-redshift clusters
($z<0.06$) using the MACHO Camera mounted on the Mount Stromlo $50''$
telescope. Comparing MSACSS and
our survey, we find that while MSACSS found more SNe (52, some $50\%$ of
which are probably cluster members), WOOTS had complete spectroscopic
follow-up observations (compared to $\le 50\%$ for MSACSS).
These additional observations allowed us to discover
and characterize the population of intergalactic SNe, presented by
Gal-Yam et al. (2003a). The fact that MSACSS targeted lower-$z$
clusters, as well as the availability of the wide-field, two-color MACHO camera,
enabled MSACSS to acquire better follow-up photometry and
measure the SN light curves (Germany et al. 2004).  
Both surveys acquired data suitable for the calculation of cluster SN
rates (Paper II), though the rates from MSACSS remain unpublished.

Mannucci et al. (2007) recently reported a measurement of the SN rate
in local ($z<0.04$) clusters, which was derived from the cluster-SN subsample
within the large dataset of events found in five historical SN surveys
targeting individual galaxies analyzed by Cappellaro et al. (1999). 
Sand et al. (2007) present preliminary results from
a survey with very similar design and goals to those of WOOTS, 
conducted with the 90'' telescopes at Steward Observatory. 
         
Moving on to higher redshifts, we have been pursuing ground-based cluster
SN surveys at $z \approx 0.2$ using numerous telescopes (e.g., Gal-Yam et
al. 2003c,d; 2005). Following a different
approach, we have measured the cluster SN rate in high-redshift clusters
using a handful of SNe discovered in archival {\it Hubble Space Telescope
(HST)} data (Gal-Yam, Maoz, \& Sharon 2002), and are in the process of
conducting an {\it HST} survey of rich $z \approx 0.6$ clusters
in order to obtain a more accurate measurement (Sharon et al. 2005; 2006).
Finally, the Supernova Cosmology Project group is leading a large {\it HST}
program to search for SNe~Ia in the most distant known galaxy clusters
(Dawson et al. 2005). These {\it HST} efforts should ultimately allow
the evolution of cluster SN rates to be traced to $z>1$. 

\section{Conclusions} 

We have conducted a survey for SNe and other transients in the fields of
rich Abell galaxy clusters at $0.06 \le z \le 0.2$. Using unfiltered imaging with
the Wise $1~$m telescope, we have achieved a sensitivity limit of $R \approx
21.5$ mag for variable sources.
An automated pipeline was written to reduce and analyze the data,
using image subtraction to detect variability.  
A dozen SNe were discovered and
spectroscopically confirmed. Seven SNe turned out to have
occurred in cluster galaxies, while five are field events. Eleven of the
SNe (including all cluster events) are apparently SNe~Ia, typically discovered near
maximum light. From the statistics of non-cluster SNe, and relying on the fact that
the WOOTS SN sample is complete (i.e., all SN candidates have been either confirmed
or ruled out), we can estimate that flux-limited SN surveys with similar limiting
magnitudes will be dominated ($\sim80\%$) by SNe Ia.  

Our follow-up spectroscopy
shows that AGNs may mimic SNe, and that reliably rejecting all AGNs
without spectroscopy is difficult.
Two of our cluster SNe had no visible hosts in the WOOTS images,
nor in deep Keck images,
and were subsequently shown to be the first good candidates for intergalactic
SNe, whose progenitor stars are not associated with galaxies
(Gal-Yam et al. 2003a). The
cluster SN sample is suitable for the calculation of the SN rate in clusters
in this redshift range, as reported in Paper II.
The survey also detected several quasars behind the cluster sample,
as well as numerous asteroids and variable stars.   

\section*{Acknowledgments} 

This project would not have come to final fruition without the diligence and
perseverance of K. Sharon. We are grateful to D. Reiss, E. O. Ofek, and
D. Poznanski for help and advice. F. Patat, M. Turatto, D. C. Leonard, A. G.
Riess, P. Leisy, O. Hainaut, T. Sekiguchi, G. Aldering, P. Nugent, C. Sorensen,
B. Schaefer, A. J. Barth, and R. Pogge provided follow-up data and observations.
Special thanks go to R. Stathakis for supporting observations
at the AAO, and to the ESO Director for Director's Discretionary observations
of several WOOTS SNe.  We acknowledge A. Howell for supplying a copy of
the {\tt Superfit} code. We also thank S. Ben Gigi, J. Dann, and Y. Lipkin
for helping with observations at Wise, and the entire Wise Observatory
staff at Mizpe Ramon and at the Tel Aviv headquarters, without whose
assistance this work would not have been possible.
Our colleagues in the Astrophysics Department at Tel Aviv
University observed WOOTS sources during their observing
runs at the Wise 1~m telescope. We are grateful to the staff at the
W. M. Keck Observatory for their assistance. The Keck Observatory
is operated as a scientific partnership among the California Institute
of Technology, the University of California, and the National Aeronautics
and Space Administration (NASA); the Observatory was made possible by
the generous financial support of the W. M. Keck Foundation.
This research has made use of the NASA/IPAC Extragalactic Database (NED),
which is operated by the Jet Propulsion Laboratory, California Institute
of Technology, under contract with the National Aeronautics and Space
Administration. A.G. acknowledges support
by NASA through Hubble Fellowship grant \#HST-HF-01158.01-A awarded by
the Space Telescope Science Institute (STScI), which is operated by AURA,
Inc., for NASA, under contract NAS 5-26555.
This work was supported by a grant from the Israel Science Foundation (D.M.).
The work of A.V.F.'s group at UC Berkeley is supported by National Science
Foundation (NSF) grant AST-0607485, as well as by NASA grant GO--10493
from STScI. P.G. acknowledges support from NSF grants AST-0307966 and AST-0607852.

\newpage 

\begin{center}
    {\bf Appendix A: The CPM Image-Subtraction Algorithm}
\end{center} 

Proper image subtraction requires matching the point-spread functions
(PSFs) of two images. The PSF varies significantly
between images due to variable atmospheric, dome, and telescope seeing,
as well as tracking errors and atmospheric dispersion (Filippenko 1982) when
observing far from the zenith. One can describe the combination of these effects
as a convolution of the undistorted source image $S$ with a distorting kernel
$D$ whose shape varies from image to image. If we denote images obtained at various
epochs by $I_i$, then for each epoch $i$, 
\begin{equation}
I_i = S \star D_i~~,
\end{equation}
where $\star$ denotes a convolution. In Fourier space, we can use the convolution
theorem and replace convolutions with products. If we denote the Fourier
transform of $X$ by $\tilde{X}$, we can define the kernel $\tilde{K_{ij}}$
\begin{equation}
\tilde{K_{ij}} \equiv {\tilde{I_j} \over \tilde{I_i}} = {\tilde{D_j} \over \tilde{D_i}} ~~,
\end{equation}
which, when applied to image $I_i$, will match its PSF to that of image $I_j$
in the following manner:
\begin{equation}
\tilde{I_i} \times \tilde{K_{ij}} = \tilde{I_i} \times {\tilde{D_j} \over \tilde{D_i}} =
S \times \tilde{D_i} \times {{\tilde{D_j}} \over {\tilde{D_i}}} =
S \times \tilde{D_j} = \tilde{I_j}~~.
\end{equation} 
Note that to calculate $K_{ij}$ we can replace the ratio of the spatial transforms
of the images ${\tilde{I_j}/\tilde{I_i}}$ whose calculation is computationally
prohibitive for large images, by the ratio of the PSFs ${\tilde{D_j}/\tilde{D_i}}$,
with each PSF measured from small subsections of the images centered on one
or more isolated, bright stars. This Fourier division method enables one to degrade
the PSF of the better image to that of the one of poorer quality (broader PSF). ISIS finds an optimal convolution kernel directly in real space assuming the kernel to be
composed of a set of basis kernels, with the coeeficients 
comupted through least-square analysis (see Lupton \&
Alard 1998 and Alard 2000 for
further details).

Our algorithm is based on a simple
approach. For each image we determine the PSF $D$ by inspecting several
bright, isolated stars. While the algorithms described above ({\tt psfmatch}
and ISIS) try to find
the optimal kernel $K$ which will degrade the PSF of the better image to match
that of the poorer image, our algorithm convolves
each image with the PSF of the second image, in analogy to the arithmetic
``common denominator.'' Since the following simple equation holds, 
\begin{equation}
I_i \star D_j = S \star D_i \star D_j = S \star D_j \star D_i = I_j \star D_i~~,
\end{equation}             
we get, in principle, two perfectly matched images. The drawback of the method
is that, as both images are degraded, the final PSF is even broader
than the PSF of the worse image. This is equivalent to a seeing degradation
of up to $\sqrt 2$ in the PSF width, if both PSFs are of similar breadth.

\begin{figure}
\includegraphics[width=145mm]{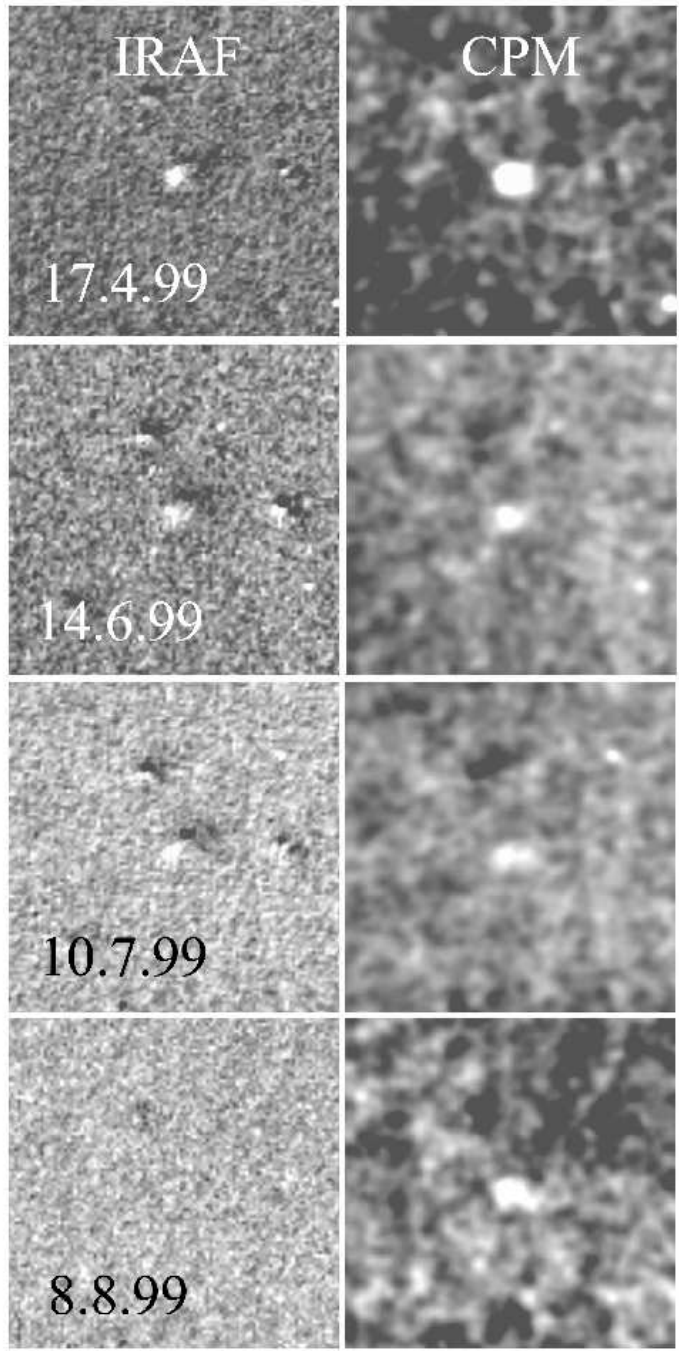}
\caption{Comparison of image subtraction using IRAF$/${\tt psfmatch}
and CPM. A template image has been subtracted from WOOTS
images of Abell 1966, showing SN 1999ay, obtained between
April and August 1999, using IRAF$/${\tt psfmatch} (left) and
CPM (right). We can see that the signal obtained using CPM subtraction
is generally stronger, and the residuals in the cores of nearby
galaxies weaker or absent. In particular, CPM allows us to
recover SN 1999ay in data from July 1999 (and marginally in August 1999),
up to five months after discovery, while IRAF$/${\tt psfmatch} subtraction
of these data shows no signal at all. The light curve of this SN~IIb is
typical of its class and is shown in Fig.~\ref{sn1999ay-lc-fig}.}
\label{figcpm}
\end{figure}

In principle, this loss limits the detection of faint variable sources, since the
signal is smeared over more pixels. However,
we find that CPM significantly reduces subtraction residuals
originating from the Fourier division performed by {\tt psfmatch}. For the WOOTS
survey, the reduction in subtraction residuals more than
compensates for the seeing degradation. An example is shown in Figure~\ref{figcpm}.
We have also compared the performance of CPM with ISIS. Here the results are
less conclusive, but usually we find that CPM works better. In addition, the
fact that the user is in full control of the operation of the software (e.g.,
the selection of PSF stars) makes CPM more tractable.  

\begin{figure}
\plotone{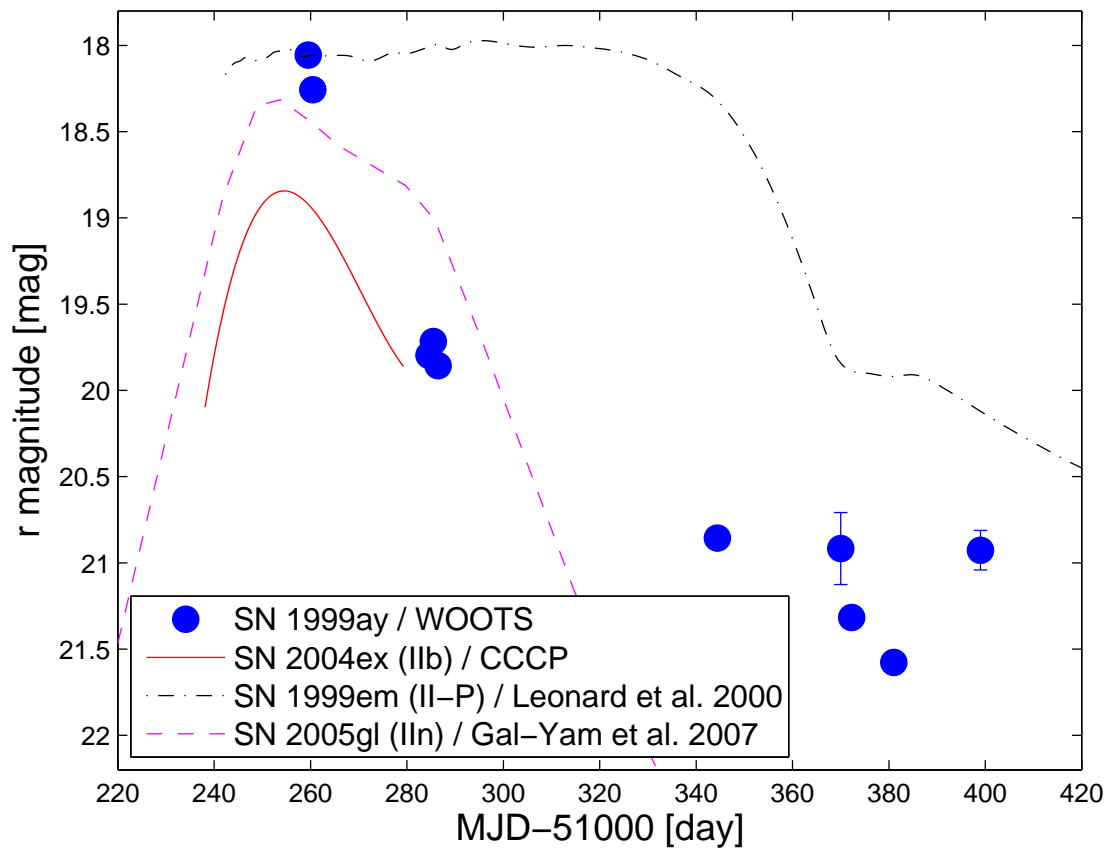}
\caption{A light curve of SN 1999ay produced using the {\tt mkdifflc}
photometry pipeline (Gal-Yam et al. 2004) which is based on CPM image
subtraction. Comparison with observations of various SNe II
shows a similarity to the light curve of the Type IIb SN 2004ex obtained
by the Caltech Core-Collapse Project (CCCP; Gal-Yam et al. 2006b, 2007),
supporting the spectroscopic identification
of this object as a Type IIb event. Unfiltered WOOTS photometry was calibrated
onto the SDSS $r$-band zeropoint using SDSS photometry of nearby stars, while the
light curves of SNe 1999em, 2004ex, and 2005gl were shifted temporally to
match the spectroscopic age of SN 1999ay (Fig.~\ref{sn1999ay-spec-fig})
and arbitrarily scaled vertically for clarity.}
\label{sn1999ay-lc-fig}
\end{figure}

This method has, since its
development, been applied in cases where accurate host-galaxy subtraction
proved critical (see, e.g., Gal-Yam et al. 2004; 2006a). In the context of the present work,
it allows us to measure the light curve of SN 1999ay (Fig.~\ref{sn1999ay-lc-fig}).

The similar shape of the light curve to that of other SNe IIb (such as
SN 2004ex, Fig.~\ref{sn1999ay-lc-fig}) supports its spectroscopic identification.
Having discovered this event probably a few days after peak magnitude, our
observations imply a lower limit on the absolute peak magnitude
of $M_{r}<-18.4$ mag. We do not correct
for host-galaxy extinction, which the spectroscopy indicates is significantly below
that of SN 1993J, for which Richmond et al. (1994) estimate $A_V=0.25-1$ mag. Galactic
extinction in this direction is negligible, and we have assumed the SDSS redshift
$z=0.044$ for the host.
The implied absolute magnitude is in the upper range of the distribution for SNe IIb
(Richardson et al. 2006; Modjaz et al. 2007). 

\newpage 

\begin{center}
    {\bf Appendix B: SNe in Random Galaxies from LOSS}
\end{center} 

In order to investigate the fraction of SNe Ia observed in a non-targeted,
relatively shallow SN survey, we have examined all the SNe discovered by
the LOSS/KAIT program (Filippenko et al. 2001; Filippenko \& Li 2008; see
also http://astro.berkeley.edu/$\sim$bait/kait.html) during the years
1999--2006. We have found 32 events
(listed in Table~\ref{losstable} below) that occurred in random galaxies
projected near LOSS target galaxies. Of these, 30/32 were spectroscopically
classified (Table~\ref{losstable}), making this among the largest and best-studied
sets of SNe discovered by a non-targeted (``blind'') SN survey (with a typical
limiting magnitude of $R \approx 19$ mag), obtained as a byproduct
of the main (targeted) LOSS program. Of the 30 events with spectroscopic types,
21 ($70\%$) are of Type Ia, consistent with our WOOTS result ($80\%$) based on a
smaller number of events from a somewhat deeper survey.  

\begin{table}[h]
\caption{SNe in Random Galaxies from LOSS}
\vspace{0.2cm}
\begin{centering}
\begin{tabular}{llll}
\hline
Supernova  & Type & Supernova & Type \\
\hline
SN 1999ce & Ia & SN 1999co & II \\
SN 2000Q  & Ia & SN 2000dd & Ia \\
SN 2001bp & Ia & SN 2001ei & Ia \\
SN 2001es & ? & SN 2002cc & Ia \\
SN 2002ct & ? & SN 2002eu & Ia \\
SN 2002ey & Ia & SN 2002hi & IIn \\
SN 2003ah & Ia & SN 2003ev & Ic \\
SN 2003go & IIn & SN 2003hw & Ia \\
SN 2003im & Ia & SN 2004U & II \\
SN 2004V & II & SN 2004X & II \\
SN 2004Y & Ia & SN 2004dz & Ia \\
SN 2005X & Ia & SN 2005ac & Ia \\
SN 2005ag & Ia & SN 2005eu & Ia \\
SN 2005kf & Ic & SN 2006bw & Ia \\
SN 2006dw & Ia & SN 2006is & Ia \\
SN 2006iu & II & SN 2006lu & Ia \\
\hline
\end{tabular}
\end{centering}
\label{losstable}
\end{table}  

\end{document}